\documentclass[12pt,amssymb]{article} %Latex2e

\usepackage{mathrsfs}
\usepackage{amssymb}
\usepackage[dvips]{graphicx}
\newcommand{\newsection}[1]{
\addtocounter{section}{1} \setcounter{equation}{0}
\setcounter{subsection}{0} \addcontentsline{toc}{section}{\protect
\numberline{\arabic{section}}{{\rm #1}}} \vglue .6cm \pagebreak[3]
\noindent{ \bf  \thesection. #1}\nopagebreak[4]\par\vskip .3cm}
\newcommand{\newsubsection}[1]{
\addtocounter{subsection}{1}\setcounter{subsubsection}{0}
\addcontentsline{toc}{subsection}{\protect
\numberline{\arabic{section}.\arabic{subsection}}{#1}} \vglue .4cm
\pagebreak[3] \noindent{\it \thesubsection.
#1}\nopagebreak[4]\par\vskip .3cm}

\makeatletter
\newcommand{\seclabel}[1]{%
  \@bsphack
  \protected@write\@auxout{}%
     {\string\newlabel{#1}{{\thesection}{\thepage}}}
  \@esphack
  }
\newcommand{\subseclabel}[1]{%
  \@bsphack
  \protected@write\@auxout{}%
     {\string\newlabel{#1}{{\thesubsection}{\thepage}}}
  \@esphack
  }
\newcommand{\tablabel}[1]{%
  \@bsphack
  \protected@write\@auxout{}%
     {\string\newlabel{#1}{{\arabic{tabnum}}{\thepage}}}
  \@esphack
  }
\makeatother

\renewcommand{\theequation}{\thesection.\arabic{equation}}

\newlength{\extraspace}
\newlength{\extraspaces}
\newcounter{dummy}
\newcommand{\bc}{\begin{center}}
\newcommand{\ec}{\end{center}}
\newcommand{\be}{\begin{equation}
\addtolength{\abovedisplayskip}{\extraspaces}
\addtolength{\belowdisplayskip}{\extraspaces}
\addtolength{\abovedisplayshortskip}{\extraspace}
\addtolength{\belowdisplayshortskip}{\extraspace}}
\newcommand{\ee}{\end{equation}}

\newcommand{\ba}{\begin{eqnarray}
\addtolength{\abovedisplayskip}{\extraspaces}
\addtolength{\belowdisplayskip}{\extraspaces}
\addtolength{\abovedisplayshortskip}{\extraspace}
\addtolength{\belowdisplayshortskip}{\extraspace}}
\newcommand{\ea}{\end{eqnarray}}

\newcommand{\ban}{\begin{eqnarray*}
\addtolength{\abovedisplayskip}{\extraspaces}
\addtolength{\belowdisplayskip}{\extraspaces}
\addtolength{\abovedisplayshortskip}{\extraspace}
\addtolength{\belowdisplayshortskip}{\extraspace}}
\newcommand{\ean}{\end{eqnarray*}}
\newcommand{\baa}{
\addtocounter{equation}{1} \setcounter{dummy}{\value{equation}}
\setcounter{equation}{0}
\renewcommand{\theequation}{\thesection.\arabic{dummy}\alph{equation}}
\begin{eqnarray}
\addtolength{\abovedisplayskip}{\extraspaces}
\addtolength{\belowdisplayskip}{\extraspaces}
\addtolength{\abovedisplayshortskip}{\extraspace}
\addtolength{\belowdisplayshortskip}{\extraspace}}
\newcommand{\eaa}{
\end{eqnarray}
\setcounter{equation}{\value{dummy}}
\renewcommand{\theequation}{\thesection.\arabic{equation}}}

\input epsf

\newcounter{fignum}
\newcounter{tabel}

\newcounter{tabnum}
\newcommand{\vev}[1]{\left\langle #1\right\rangle}
\newcommand{\ket}[1]{\left| #1 \right\rangle}

\newcommand{\half}{\frac{1}{2}}
\newcommand{\del}{\partial}
\newcommand{\delb}{\bar{\del}}
\newcommand{\eol}{\nonumber \\}

\newcommand{\cO}{{\cal O}}
\newcommand{\Ext}{{\rm Ext}}
\newcommand{\Hom}{{\rm Hom}}

\newcommand{\Q}{{\cal Q}}

\newcommand{\V}{{\cal V}}
\newcommand{\M}{{\cal M}}
\newcommand{\D}{{\cal D}}

\newcommand{\bt}{{\bf 10}}
\newcommand{\bfv}{{\bf 5}}
\newcommand{\bfb}{{\overline{\bf 5 \!}\,}}
\newcommand{\btb}{{\overline{\bf 10 \!}\,}}

\begin{document}

%%%%%%%%%%%%%%%%%%%%%%%%%%%%
%
% Date and preprint number
%
%%%%%%%%%%%%%%%%%%%%%%%%%%%%%

\begin{flushright}
May 2010\\
%{\tt hep-th/yymmnnn}\\
\end{flushright}
\vspace{2cm}

\thispagestyle{empty}

%%%%%%%%%%%%%%%%%%%%%%
%
% Title & abstract
%
%%%%%%%%%%%%%%%%%%%%%%%

\begin{center}
{\Large\bf  MSW Instantons
%Phenomenology
 \\[13mm] }

{\sc Ron Donagi}\\[2.5mm]
{\it Department of Mathematics, University of Pennsylvania \\
Philadelphia, PA 19104-6395, USA}\\[9mm]

{\sc Martijn Wijnholt}\\[2.5mm]
{\it Physics Department, Harvard University\\
17 Oxford Street \\
Cambridge MA 02138, USA }\\
[25mm]

 {\sc Abstract}

\end{center}

We analyze $M5$-instantons in $F$-theory, or equivalently
$D3$-instantons with varying axio-dilaton, in the presence of
7-brane gauge groups. The chiral two-form on the $M5$-brane plays an
important role, because it couples the $M5$ brane to vector
multiplets and charged chiral fields. The chiral two-form does not
have a semi-classical description. However if the worldvolume of the
$M5$ admits a fibration over a curve with surface fibers, then we can reduce the worldvolume
theory to an `MSW' CFT by shrinking the surface. For this class of
MSW instantons, we can use heterotic methods to do computations.
We explain this in some detail using the physical gauge approach. We
further compare $M5$-instantons with $D3$-instantons in perturbative
type IIb and find some striking differences. In particular, we show
that instanton zero modes tend to disappear and constraints from
chirality on instanton contributions to the superpotential evaporate
for finite string coupling.

\newpage

\renewcommand{\Large}{\normalsize}

\tableofcontents

\newpage

\newsection{The $M5$ instanton in $F$-theory}

The superpotential in an $F$-theory compactification is independent
of the K\"ahler moduli, to all orders in large volume perturbation
theory. However, dependence on the K\"ahler moduli may arise
non-perturbatively. As the leading corrections to the
superpotential, instanton corrections are important for a number of
issues in phenomenological models. For instance we may use them for
generating scales, like the scale of supersymmetry breaking, or for
lifting flat directions. In $F$-theory models, the natural objects
generating such corrections are $D3$-instantons, as they remain
$D3$-instantons under $SL(2,{\bf Z})$ transformations. Other
instantons, such as worldsheet/$D1$-instantons of perturbative type
IIb, are normally absent in $F$-theory due to $SL(2,{\bf Z})$
monodromies.

$F$-theory vacua are essentially supergravity backgrounds. Rules for
computing corrections due to Euclidean solitons in supergravity have
been proposed by \cite{Becker:1995kb,Harvey:1999as,Witten:1999eg}.
The basic idea is to compute a correlator of vertex operators in the
twisted worldvolume theory of the instanton. Since the worldvolume
theory of a $D3$-brane is the maximally supersymmetric Yang-Mills
theory, naively this means we should compute correlators in this
Yang-Mills theory, or perhaps some twisted version thereof.

However it is not really correct to conclude that $SL(2,{\bf Z})$
monodromies do not affect the worldvolume theory. The coupling
constant of the Yang-Mills theory is the axio-dilaton, which varies
from point to point over the worldvolume and is usually
multi-valued, because the discriminant locus typically intersects
the worldvolume. This means that generically the worldvolume theory
cannot be discussed in terms of a weakly coupled Yang-Mills theory
with only electric degrees of freedom, as circling around a branch
cut turns electric into magnetic degrees of freedom.
%Alice string

As always in $F$-theory, the way to deal with the branch cuts is to
switch to different variables. A $D3$-instanton contribution to the
superpotential is of the form
\be \Delta W\ \sim\ f(\Phi)\, e^{-T} \ee
where $T$ is a K\"ahler modulus and $\Phi$ denotes complex
structure moduli. Thus by varying K\"ahler moduli we may
extrapolate to another regime without changing the pre-factor
$f(\Phi)$. The most common way to define an $F$-theory vacuum
\cite{Vafa:1996xn} is by taking a limit of an $M$-theory vacuum on
an elliptically fibered Calabi-Yau four-fold with $G_4$-flux.
Essentially this entails a change of variable from the axio-dilaton
to the coefficients of the Weierstrass equation, which are defined
globally without branch cuts. The $D3$ instanton with varying
coupling constant in $F$-theory lifts to an $M5$ instanton in
$M$-theory, wrapping the elliptic fiber. Therefore a computation on
a $D3$-brane with varying axio-dilaton can be reformulated as a
computation in the $M5$ worldvolume theory. The $M5$-brane theory
describes both electric and magnetic degrees of freedom, and is not
weakly coupled.

The $M5$ worldvolume theory is the six-dimensional $(0,2)$-theory,
with five scalars $\phi$, a self-dual two-form $B^+$, and sixteen
superpartner fermions $\theta$. Unlike a $D3$-brane in perturbative
IIb, it is a chiral theory. It must have somewhat exotic
properties, as expected from the above observations, and this
complicates the computations. Up to some universal zero modes, the
contribution to the superpotential is given by the partition
function, which is of the schematic form
\be Z_{M5} \ = \ Z_{\phi}\, Z_\theta\, Z_{B^+} \ee
The partition functions of the fermions and scalars are relatively
straightforward. The fermion kinetic terms on the $M5$ are
schematically of the form
\be \int_{M5} d^6x\, \theta(\not \!\!D + \not \!\!G) \theta \ee
As we will briefly review later, the analysis of fermion zero modes
is straightforward in the absence of $G$-flux, and has been
addressed even in backgrounds with $G$-flux. Although global
$G$-fluxes are hard to work with, one can easily find examples where
they don't affect the fermionic zero modes. The (twisted) scalars
are also manageable.

The partition function of the chiral two-form $Z_{B^+}$ is a
slightly more complicated story. The chiral two-form acts as the
bookkeeping device that allows us to keep track of the electric and
magnetic gauge fields on the $D3$-brane together with the
monodromies. It does not admit a conventional Lagrangian
description, but its partition function can be related to a theta
function on the intermediate Jacobian ${\cal J}(M5)$ through
holomorphic factorization \cite{Witten:1996hc}, and thus in
principle we can compute it.
%\footnote{The anomaly does not appear
%however if $h^{0,3}=0$, i.e for an isolated instanton.}

So it seems there's a fairly coherent story for $M5$-instantons in
$F$-theory. Nevertheless in light of the recent advances regarding
model building in $F$-theory
\cite{Donagi:2008ca,Beasley:2008dc,Hayashi:2008ba}, it has become
clear that the current literature is insufficient and that a number
of important general aspects of $M5$-instantons have never been
discussed. Most of all one would like to understand how the
instanton interacts with charged degrees of freedom. This will be
the main subject of the present paper.

In section \ref{M5Instanton} we will see how the $M5$-instanton can
contribute to couplings of holomorphic fields, including couplings
which are forbidden in perturbation theory by gauged $U(1)$
symmetries, generalizing analogous stories in the heterotic string
and perturbative type IIb. This is one of the main new results.
Along the way we will clarify a few issues regarding $M5$ zero
modes. However explicitly computing the partition function in terms
of the compactification data is not straightforward, and involves
transcendental methods rather than algebraic ones. In general it
also appears to suffer from a holomorphic anomaly.

In order to test our general understanding, we are going to compare
with certain limits. In section \ref{IIbLimit} we will consider the
Sen limit \cite{Sen:1997gv,Clingher:2012rg}. Note that the Sen
limit involves changing the complex structure moduli, so the
pre-factor $f(\Phi)$ will not stay the same. In the Sen limit the
string scale is parametrically small compared to the $10d$ Planck
scale, and one could use the method of Ganor strings
\cite{Ganor:1996pe} to derive the collective coordinates of the
instanton. We find a number of new and interesting results. For
example we argue that the partition function $Z_{B^+}$ reproduces
not only the partition function $Z_F$ for the gauge field on the
$D3$-instanton, but also the partition function $Z_{\lambda_{37}}$
of the $D3$-$D7$ Ganor strings. (This may be shown more precisely
using the recent results of \cite{Clingher:2012rg}). This is just
as well since it would not make sense to quantize strings at finite
coupling, and our result says that the correct generalization of
$Z_{\lambda_{37}}$ to finite string coupling involves the chiral
two-form $B^+$.

But the Sen limit certainly does not capture the full story.
Essentially it imposes an $SO(2n)$ structure and therefore cannot
properly describe the exceptional structures that play an important
role in the phenomenological models. Indeed if one takes a model
with exceptional structures, then one encounters singularities with
non-perturbative physics in the Sen limit \cite{Donagi:2009ra}. As
we will see in section \ref{IIbLimit}, another way the $SO(2n)$
structure manifests itself is that one gets certain extra light
$U(1)$ symmetries in the Sen limit which are absent for generic
$F$-theory models. These extra $U(1)$ symmetries impose stringent
constraints. For example, it was realized a number or years ago
that the extra $U(1)$ symmetries lead to an apparent conflict
between the presence of chiral matter and instanton contributions
to the superpotential in perturbative type IIb that help stabilize
the K\"ahler moduli \cite{Blumenhagen:2007sm}. One of the main
points that we will make here is that these `extra' $U(1)$
selection rules are artefacts of the $g_s \to 0$ limit and that
these constraints are absent in more generic $F$-theory models.

The second limit we will consider is the limit in which one splits
off a $dP_9$-fibration \cite{Donagi:2012ts}. This is also a limit
in complex structure moduli space, so again the pre-factor
$f(\Phi)$ will not stay the same. However in contrast to the Sen
limit, this limit does not touch the elliptic fibration over the
GUT divisor, and in particular it preserves the exceptional
structures. So this limit should capture the generic features of
$F$-theory models, at least as far as the coupling to charged
degrees of freedom is concerned.

In this limit we have the following situation: the $M5$-brane
splits into two pieces, one of which (namely the part that talks to
the local model) is wrapping a $dP_9$-surface. Thus we are
interested in the situation where besides the elliptic fibration,
the $M5$-brane also admits a surface fibration over a complex curve
$\Q$:
\be
\begin{array}{ccc}
  S & \to & M5\\
   &  & \downarrow \\
   &  & \Q
\end{array}
\ee
Now recall that the pre-factor $f(\Phi)$ depends only on complex
structure moduli, and is independent of the size of $S$. Then we
may as well do the computations in the limit that $S$ shrinks to
zero. In this limit, the $M5$ collapses to a string, and the
worldvolume theory of the $M5$ reduces to an effective ``MSW'' CFT
\cite{Maldacena:1997de} on $\Q$. (More mathematically, one gets a
natural relation between the partition functions using a `cylinder
map.') The resulting theory is similar to the $E_8 \times E_8$
heterotic string, except we only have a single $E_8$ in our case.
The chiral two-form reduces to chiral right- or left-moving
scalars, which may be fermionized. The partition function is
identified with a section of a determinant line bundle, and its
zeroes can be understood as certain Dolbeault cohomology groups.
Therefore in this class of examples we can be pretty explicit,
while still retaining the essential features of $M5$-instantons,
like a varying axio-dilaton with branch cuts on the worldvolume.
This is the subject of section \ref{HeteroticInstanton}.

Having mapped the problem to a heterotic-like set-up, one could try
to apply heterotic techniques to compute instanton correlators.
From the $F$-theory perspective it is natural to try and apply the
rules of \cite{Becker:1995kb} et al to compute instanton
correlators, instead of using the old worldsheet techniques. The
MSW string is generally not fundamental and so the old worldsheet
methods should not be necessary. In fact such an approach was
already proposed by Witten \cite{Witten:1999eg} quite a while ago.
We will elaborate on this idea and see how it can be used to
compute instanton correlators of charged fields.

Moreover in this set-up one can do very explicit and algebraic
computations. We will explain how the heterotic computations of
\cite{Buchbinder:2002ic} may be reinterpreted as calculating (a
piece of) the partition function $Z_{B^+}$ in our limit.

In section \ref{IIbLimit} we include a discussion of reducible
brane configurations, which applies more broadly than the context
encountered here. We have elaborated on this material in
\cite{Donagi:2011jy,Donagi:2011dv}.

 A number of results on $D3$-instantons in type IIb and $T$-dual set-ups
have appeared in recent years. See \cite{Blumenhagen:2009qh} for an
extensive review of the literature on $D$-instantons in
perturbative type II theories, and \cite{Heckman:2008es} for
configurations close to those discussed here. Our discussion is
meant to address the interaction of $D3$-instantons with mutually
non-local 7-branes at finite string coupling. We will see how some
of the old results are recovered as $g_s \to 0$.

Several related works appeared recently, see
\cite{Cvetic:2009ah,Cvetic:2010rq} and \cite{Blumenhagen:2010ja}.
Our `heterotic' approach to $M5$-instantons and a number of our
results were announced at the workshop ``GUTS and Strings'' in
Munich \cite{Munich}. We would also like to point out a new and
precise description of the Sen limit \cite{Clingher:2012rg}, which
allows one to prove some of the claims made in section
\ref{IIbLimit}.

\newpage

\newsection{Instantons in the heterotic string}
\seclabel{HeteroticInstanton}

\newsubsection{Green-Schwarz approach}

In the Green-Schwarz formalism, the worldsheet instanton amplitudes
look quite similar to those of $D$-instantons, so one may try to use
$D$-instanton inspired techniques to calculate worldsheet instanton
effects. In \cite{Witten:1999eg} this is called the physical gauge
or Green-Schwarz approach.

Consider a curve $\Q$ in the heterotic Calabi-Yau $Z$ with bundle
$V$. The physical degrees of freedom living on $\Q$ are as follows.
The left-movers form an $SO(32)$ or $E_8 \times E_8$ current
algebra. We will use the fermionic formulation. Then the left-moving
fermions live in
\be \lambda\ \in\ \Gamma({\Q},{\cal V}(-1)|_\Q) \ee
The structure group of $\V$ is the manifest symmetry group that is
visible in the fermionic formulation. For the $SO(32)$ heterotic
string, $\V$ is an $SO(32)$ bundle. For the $E_8 \times E_8$
heterotic string, which is the case we will be interested in here,
$\V$ denotes an $SO(16)\times SO(16)$ bundle. The fermions satisfy
a Majorana condition in Minkowki signature, but we have to complexify
them when we work in Euclidean signature.

For the right-movers, we use GS variables $(X,\Theta)$ and restrict
to physical gauge, fixing world-sheet reparametrizations by setting
the longitudinal coordinates $X^{//}(z) = z$ and using kappa symmetry
to set $P_+\Theta =
0$. Then the remaining right-moving bosons live in
\be X^\perp \ \in \ \Gamma(\Q,\bar \cO\oplus \bar \cO \oplus \bar N) \ee
and the right-moving fermions, collectively called $\Theta^\perp$, live in
\ba
 \chi_{\dot \alpha}^{\bar m} &\in& \Gamma(\Q, \bar N|_\Q) \otimes S^- \eol
 \theta^\alpha &\in &  \Gamma(\Q, \bar \cO_\Q)\otimes S^+ = {\bf C}^2 \eol
 \theta^\alpha_{\bar z} &\in & \Gamma(\Q, \bar \Omega^1_\Q)\otimes S^+
 \ea
We have adopted the notation of \cite{Beasley:2005iu}, where the
bar indicates right-moving fields which we take to be
anti-holomorphic. The zero modes are given by the global
holomorphic sections for the left-movers, or the global
anti-holomorphic sections for the right-movers. At the least we
have four bosonic zero modes, for the four translation symmetries
broken by the instanton, and two right-moving fermionic zero modes.

The fermion bilinears in the $4d$ supergravity action are given by
\cite{Wess:1992cp}
\be
e^{K/2} [W \psi \gamma \psi + D_iW \chi^i \gamma \psi + D_iD_jW \chi^i \chi^j ]
\ee
The contribution of an instanton wrapped on $\Q$ to the
superpotential can be obtained by computing a correction to the
$\vev{\chi\chi}$ correlator. The end result is surprisingly simple;
the contribution to the superpotential is simply given by the
world-volume partition function, obtained by integrating out all
the physical degrees of freedom on $\Q$:
\be\label{PF}  Z[A,g,B] = \int dX^\perp\, d\Theta^\perp\, d\lambda \, e^{-S_\Q[A,g,B]}
\ee
It is convenient to factor out the universal zero modes:
\be \Delta S_{4d}  = \int d^4 x d^2 \theta \, \Delta W, \qquad \Delta W =
\int d\hat X\, d\hat \Theta\, d\lambda\, e^{-S_\Q[A,g,B]} \ee
We get a non-zero contribution only when $\Q$ is a rational curve.
Higher genus curves would carry additional $\theta^\alpha_{\bar z}$
zero modes, and integrating over them leads to multi-fermion
$F$-terms but not superpotential contributions.

Let us further assume that our instanton is isolated, i.e. it has
only the universal zero modes and no more. As usual in
supersymmetric theories, we only need to evaluate the classical
action and the one-loop determinant around the classical solution:
\be\label{InstantonFormula} \Delta W\ =\ {{\rm Pfaff}' (D_F)\over
\sqrt{ {\det}' (D_B)}}\, e^{-T_\Q} \ee
Here we defined the K\"ahler modulus as
\be T_\Q\ =\ \int_\Q J - {i} \int_\Q B \ee
The Pfaffian in (\ref{InstantonFormula}) arises because we
complexified the fermions in Euclidean space, so we must take a
square root to factor out the extra contributions. A prime denotes
omission of zero modes. Specializing this to our isolated instanton,
we get the following contribution to the space-time superpotential
\cite{Witten:1999eg}:
\be\label{HetPartition} \Delta W \sim  {{ P\!f} (\delb_{\V(-1)})
\over \left(\det \delb_{\cO(-1)}\right)^2 \left(\det ' \delb_\cO
\right)^2} \, e^{-T_\Q}\ee
Here ${\cal V}$ is assumed to be an $SO(32)$ bundle. For $E_8\times
E_8$, the natural guess is to replace the determinant of
$\delb_{\V(-1)}$ by the partition function of level one $E_8\times
E_8$ Kac-Moody algebra. The partition function depends on a choice
of (perturbative) vacuum, i.e. a choice of background metric,
$B$-field and gauge field.

\newsubsection{Calculating the Pfaffian}

\subseclabel{PfaffZero}

The denominator in (\ref{HetPartition}) can be computed easily but
does not depend on massless chiral fields. Let us therefore focus
on the partition function of the left-movers $Z_\lambda$, the
Pfaffian, which does depend on them. As a partition function of
fermions, the Pfaffian suffers from an anomaly:
\be Z_\lambda[A + d\varepsilon] \ = \ e^{{in\over 4\pi} \int_\Sigma
\varepsilon F} Z_\lambda[A] \ee
where $n$ is the number of species. Therefore the value of the
Pfaffian at a given point on the moduli space does not make sense.
The Pfaffian is not a function, but rather a section of a certain
line bundle over the moduli space.

Nevertheless, there is invariant information in the Pfaffian.
Although the value of a section of a line bundle is not
well-defined, the zero locus of this section is well-defined, in
fact it determines the section up to rescaling. Therefore one can
still compute the moduli dependence of the superpotential up to an
overall scalar. This strategy was successfully carried out in
several examples by \cite{Buchbinder:2002ic} and further in
\cite{Curio:2009wn}.

Let us consider the zero locus of the Pfaffian. It vanishes when the
moduli are such that $\delb_{\V(-1)}$ develops zero modes, i.e. when
$H^0(\Q,{\V}(-1)|_\Q)$ is non-trivial. Any bundle on ${\bf P}^1$
decomposes as a sum of line bundles, so we may write
\be \V|_\Q = \bigoplus_i \cO(a_i) \oplus \cO(-a_i) \ee
(The $a_i$ come in $\pm$ pairs since our $\V$ is assumed to be orthogonal.)
Using the well-known formula for the Dolbeault cohomology of line
bundles on ${\bf P}^1$, we find that
\be H^0(\Q,\V(-1)|_\Q) \ = \  \sum |a_i| \ee Hence left-moving zero
modes are absent if and only if all $a_i$ vanish, i.e. $\V|_\Q$ is a
trivial bundle \cite{Distler:1988ms}. This argument also applies to
$E_8\times E_8$ bundles, as long as the holonomy is contained in
$SO(16) \times SO(16)$.

We assume that $Z$ is elliptically fibered with section
$\sigma_{B_2}$, and $V$ is constructed through a spectral cover
$(C,L)$. We consider only rational curves that are contained in the zero section
$\sigma_{B_2}$. For these we have
\be V|_\Q \ = \  \pi_{C*}L|_\Q , \qquad V(-1)|_\Q \ = \
\pi_{C*}L(-F)|_\Q\ee
where $F$ is the class of the elliptic fiber inside the surface
$\pi_Z^{-1}(\Q)$. (One complication that arises for more general
rational curves $Q$ that are not necessarily contained in
$\sigma_{B_2}$ is that the direct image above has to be replaced by
a Fourier-Mukai transform. A more significant issue is that quite
often rigid curves $Q$ that are contained in $\sigma_{B_2}$ are the
unique effective representatives, in the threefold $Z$, of their
cohomology class, while more general rigid curves $Q$ that are not
contained in $\sigma_{B_2}$ often come in large collections that all
represent the same cohomology class in $Z$.) In our case, let us
define the spectral curve $\Sigma_{37}$ to be given by
\be \Sigma_{37}= \pi_Z^{-1}(\Q) \cap C \ee
We will explain the reason for this notation in section
\ref{IIbLimit}. By applying the direct image, we find that
\ba H^0(\Q,V(-1)|_\Q) &=& H^0(\Sigma_{37},L(-F)|_{\Sigma_{37}}), \eol
H^0(\Q,V^*(-1)|_\Q) &=& H^1(\Sigma_{37},L(-F)|_{\Sigma_{37}})^* \ea
Note that since $c_1(V) = 0$, from the Riemann-Roch formula on $\Q$
it follows that the ranks of these two Dolbeault cohomology groups
must be equal, so fermion zero modes always come in pairs. From
Riemann-Roch on $\Sigma_{37}$ we see that $L(-F)$ is a line bundle
of degree $g-1$, where $g$ is the genus of $\Sigma_{37}$. The
vanishing locus of the Pfaffian corresponds to the locus where the
Dolbeault cohomology groups on $\Sigma_{37}$ valued in the line
bundle $L(-F)|_{\Sigma_{37}}$ are non-zero.

The fact that $L$ has vanishing holomorphic Euler characteristic
(i.e. degree $g-1$), rather than vanishing degree, is crucial also
for the interpretation of the Pfaffian via theta functions. In our
set-up varying the bundle $V$ results in variation of the complex
structure of $\Sigma_{37}$ as well as the induced line bundle on
it. The line bundle $L$ on $C$ typically has no moduli, but by
varying the spectral cover, the restriction of $L$ to $\Sigma_{37}$
may vary. Hence we are dealing with a map
\be \M(C,L) \to {\cal U},  \ee
from the space $\M(C,L)$ parametrizing the pairs $C,L$ to a
"universal Picard variety" $\cal U$ which fibers over the moduli
space $\M_g$ of genus $g$ Riemann surfaces, the fiber over $C$
being the torus $Pic^{g-1}(C)$ parametrizing all complex line
bundles of degree $g-1$ on $C$. When we thus allow $\tau$ (or
equivalently, the period matrix) to vary, it becomes important that
the Pfaffian really computes $\theta/\eta$, with the $1/\eta$ being
the contribution of the massive modes. Here $\theta$ becomes a
function of $z$ and $\tau$, while $\eta$ is a function of $\tau$
alone. The vanishing locus of the Pfaffian corresponds to the
inverse image of the theta divisor, and $\eta$ becomes part of the
undetermined scaling.

The algebraic approach is based on determinant line bundles,
following Grothendieck-Knudsen-Mumford and Deligne. It is known to
be equivalent \cite{BGS3} to the transcendental approach outlined
above, but is often much more amenable to explicit calculations.
The algebraic description of the Pfaffian also makes it clear that
we get a polynomial on $\M(C,L)$.

In fact we can use the algebraic description to calculate the moduli
dependence of the Pfaffian directly. Suppose we have a family of
Riemann surfaces parametrized by $s\in S$. In the algebraic
description the determinant line bundle over $S$ is defined as
\be {\mathscr Det}\ =\ {\rm det}(H^0(\Sigma_s,L_s))^{-1}\, {\rm
det}(H^1(\Sigma_s,L_s)) \ee
Now suppose that we can establish an exact sequence of the form
\be
0\ \to\ H^0(\Sigma_s,L_s)\ \to\ W_1\ \stackrel{f}{\to}\ W_2\ \to\ H^1(\Sigma_s,L_s)\ \to\ 0
\ee
or some variation thereof. Then by general properties of determinant lines,
we get an isomorphism
\be {\mathscr Det}\ =\ {\rm det}(W_1)^{-1}\, {\rm det}(W_2)) \ee
and a canonical section $\det f$. Since the algebraic and analytic
approaches agree, we may take
\be Pf\ =\ \det \delb\ =\ \det f \ee

The authors of \cite{Buchbinder:2002ic} arrive at
the following exact sequence
\be 0\ \to\ H^0(\Sigma_{37},L(-F))\ \to\ W_1\ \stackrel{f}{\to}\ W_2\ \to\
H^1(\Sigma_{37},L(-F))\ \to\ 0
\ee
with
\be W_1 = H^1(\pi^{-1}\Q, L(-F-\Sigma_{37})), \qquad W_2 = H^1(\pi^{-1}\Q, L(-F)) \ee
Here $f$ is multiplication by the section which vanishes at
$\Sigma_{37}$, i.e. the equation of $\Sigma_{37}$ in $\pi^{-1}\Q$,
and the last map is restriction to $\Sigma_{37}$. Calculating $f$
by pushing down to $\Q$, one finds explicit formulae for the
Pfaffian as a function of the moduli \cite{Buchbinder:2002ic}.

Let us try to restate this in a more physical and less mathematical
manner. It will be useful to go through this because we get very
similar objects as in the general $M5$-brane story, but in a more
familiar setting. We have some left-moving fermions on $\Q$,
coupled to a gauge bundle, and we want to compute their partition
function (or at least the moduli dependence). Since
$V|_{\Sigma_{B_2}} = \pi_{C*}L$, we can locally think of $V$ as $n$
copies of $L$, where $n$ is the degree of the spectral cover, but
then the $n$ fermions coupled to each copy will have branch cuts.
We can eliminate the branch cuts by thinking of the $n$ fermions as
a single fermion living on a $n$-fold cover of $\Q$, which is what
we have called $\Sigma_{37}$, and coupled to $L$. Therefore we want
to compute the partition function of a chiral fermion on
$\Sigma_{37}$ coupled to $L$:
\be
 Z_\lambda[A]\ \propto\ \int d\lambda \, e^{-\int_{\Sigma_{37}} d^2 z \sqrt{g}\, \lambda \delb_A \lambda}
\ee
as well as a second fermion on $\Sigma_{37}$, obtained by lifting
the fermion on $\Q$ that is coupled to $V^*$. The line bundle should
be flat in order for the partition function to make sense. In the
fermionic description this arises because if the line bundle is not
flat, then there are always chiral fermion zero modes which can only
be absorbed by extra insertions in the path integral. In the
bosonized description, this arises because the interaction term may
be rewritten as
\be \int_\Sigma d^2 z\,\del_z \phi\, A_{\bar z}\ =\ - \int_\Sigma
d^2 z\, \phi F_{z\bar z} \ee
which implies that there is a tadpole for the chiral boson. The line
bundle $L(-F)$ that we found above actually has degree $g-1$, but we
should morally think of it as $\tilde L \otimes
K_{\Sigma_{37}}^{1/2}$ where $\tilde L$ is flat.

Partition functions of chiral fermions on a Riemann surface are
well-known objects, and we could take several points of view. Our
interest here is in the fact that up to a non-vanishing factor, they
are given by theta-functions with characteristics:
\be Z_\lambda \ \propto \ \Theta[{}^\alpha_\beta](\tau|\nu) \ee
For instance, on a genus one Riemann surface we have the well-known
expressions
\be Z_\lambda \ = \ q^{-1/24+\half \alpha^2}e^{2\pi i
\alpha\beta}\prod (1 + e^{2\pi i\beta} q^{m-\half +\alpha} z)(1+
e^{-2\pi i\beta}q^{m-\half-\alpha}z^{-1}) =
{\Theta[{}^\alpha_\beta](\tau|\nu)\over \eta(\tau)} \ee
with $q = \exp(2\pi i\tau)$ and $z = \exp(2\pi i \nu)$. This
generalizes to higher genus Riemann surfaces and also to
$M5$-branes. Let us review some aspects of the theory of
theta-functions.

We fix a basis $\{A_i, B^j\}$ of one-cycles on $\Sigma_{37}$, with
the following intersection properties:
\be A^i \cap A^j\ =\ B_i \cap B_j\ =\ 0, \quad A^i \cap B_j\ =\ \delta^i_j
\ee
Let us also fix a basis $\omega^j$ of holomorphic one-forms with the property
\be \int_{A^i}\omega_j\ =\ \delta^i_j \ee
The period matrix is defined as
\be \tau_{ij}\ =\ \int_{B_i} \omega_j \ee
Now let us fix a base point $p_0$ on $\Sigma_{37}$. Then we get a
multi-valued map from $\Sigma$ to ${\bf C}^g$ by
\be p\ \to\  \int_{p_0}^p \omega_j \ee
To get a single valued map on $\Sigma$, we have to make the identifications
\be \vec{x}\ \sim\ \vec{x} + (\vec{a} + \vec{b} \cdot \tau), \qquad
\vec{a},\vec{b} \in {\bf Z}^g \ee
The periodic identifications by the lattice $\Lambda = {\bf Z}^g +
{\bf Z}^g \tau$ make ${\bf C}^g$ into a torus, which is called the
Jacobian of $\Sigma$:
\be {\cal J}(\Sigma)\ =\ {\bf C}^g/\Lambda \ee
Said differently, a closed one-form $A$ on $\Sigma$ defines a point
on $H^1(\Sigma,{\bf R})/H^1(\Sigma,{\bf Z})\sim T^{2g}$. The Hodge
$*$-operator satisfies $*^2 = -1$, so it defines a complex structure
on $H^1$:
\be
*A^{1,0}\ = \ +i A^{1,0}, \qquad * A^{0,1}\ = \ -i A^{0,1}
\ee
The induced a complex structure on $T^{2g}$ gives us ${\cal
J}(\Sigma)$.

The Jacobian naturally comes with several additional structures.
Given a metric on $\Sigma$, we get a translationally invariant
metric on ${\cal J}(\Sigma)$ given by
\be g_{\cal J}(A,A) \ = \ \int_\Sigma A \wedge * A \ee
The associated K\"ahler form
\be \omega(A,A') \ = \ \int_\Sigma A \wedge A' \ee
defines a `principle polarization' on ${\cal J}$, i.e. it is a
symplectic form in $H^2({\cal J},{\bf Z})$ such that ${\cal
J}(\Sigma)$ has volume equal to one.

Any holomorphic line bundle on $\Sigma_{37}$ is determined by the
flux (i.e. its first Chern class), and its continuous moduli. When
the first Chern class vanishes, the continuous moduli are the
Wilson lines, i.e. the periods of a holomorphic connection. Let us
assume that the flux vanishes, so that the fermion partition
function is defined. Then the connection is a closed form, and the
Wilson lines determine a unique point on the Jacobian; the
identifications by $\Lambda$ are due to the large gauge
transformations. So we may think of the Jacobian as the moduli
space of flat connections on $\Sigma_{37}$. Therefore the partition
function naturally `lives' on ${\cal J}(\Sigma)$.

Now recall that the partition function of a chiral fermion on
$\Sigma_{37}$ is not a function on the Jacobian but a section of a
line bundle ${\cal L}$. The curvature of this line bundle is in
fact $2\pi \omega$. From the index theorem on ${\cal J}(\Sigma)$
and positivity of $\omega$, it follows that ${\cal L}$ has exactly
one section, which we want to identify with $Z_\psi$. But to fix
${\cal L}$ we also need to specify the continuous moduli.

To do this we first consider the moduli space of degree $g-1$ line
bundles. Here we have a canonical section, namely the Riemann theta
function. Its zero set, called the theta-divisor, corresponds to
degree $g-1$ line bundles with a section. Its translates by a spin
structure yields the theta-functions with characteristics. They have
the right vanishing behaviour \cite{AlvarezGaume:1986es}. Since for
each spin structure we had a line bundle with a unique section, the
partition function is uniquely determined to be proportional to the
associated theta-function with characteristics on the Jacobian.

In our case actually we are not given $\tilde L_{flat}$ and
$K_\Sigma^{1/2}$ separately, but rather the tensor product $\tilde
L_{flat}\otimes K_\Sigma^{1/2} = L(-F)$ which is defined
unambiguously. Since $L(-F)$ has degree $g-1$, as discussed above
there is actually a unique choice of theta-function, the Riemann
theta-function, which we interpret as the partition function of our
fermions. The theta-function vanishes along a divisor on ${\cal
J}(\Sigma_{37})$, the theta-divisor, which corresponds to the locus
where the line bundle $L(-F)|_{\Sigma_{37}}$ has a section, i.e.
where we get fermion zero modes.

More conceptually, our fermions are coupled to a (reducible)
$SO(2n)$ bundle, so our spectral curve is actually $\Sigma =
\Sigma_{37} \cup \rho^*\Sigma_{37}$, where $\rho$ is the
`orientifold' involution. The partition function corresponds to the
theta function of the Prym of $\Sigma$, i.e. the anti-symmetric
part of the Jacobian of $\Sigma$. This explains some of the
peculiar shifts that we saw above.

\newsubsection{Coupling to supergravity}

Although the partition function for chiral fermions suffered from a
gauge anomaly, the combined partition function, including all the
worldsheet fields and using a proper definition of the $B$-field,
must be anomaly free. So it seems that modulo possible
$R$-anomalies, the superpotential should be an ordinary function on
the moduli space. However when we couple to supergravity, in order
for the action to be even classically invariant under K\"ahler
transformations, we also need accompany the K\"ahler transformations
with a certain local $U(1)_R$ transformation. As a result of this,
when $M_{Pl}$ is finite, the superpotential is still a section of a
non-trivial line bundle over the moduli space. Let us briefly sketch
some of the structure of $N=1$ supergravity Lagrangians that lead to
this conclusion (see eg. \cite{Weinberg:2000cr}).

We can couple the globally supersymmetric Lagrangian for the matter
fields to the linearized supergravity multiplet through the Noether
currents with coupling $1/M_{Pl}$. This action is invariant under
local supersymmetry up to terms of order $1/M_{Pl}$. We then add
further terms to the Lagrangian and supersymmetry variations so that
supersymmetry is preserved to this order, and so on. One ends up
with a Lagrangian in which the fermion kinetic terms contain the
following covariant derivative:
\be D_i = \nabla_i + \half q\,\kappa^2\, \del_i K, \qquad \kappa^2 =
{1\over M_{Pl}^2} \ee
Here $\nabla$ is the ordinary covariant derivative including
K\"ahler Christoffel symbols for the non-trivial metric on the sigma
model, and $q$ is $-1$ for chiral fermions and $+1$ for gauginos and
the gravitino.

Now under a K\"ahler transformation, we have
\be \kappa^2\,K(z,\bar z) \to \kappa^2 \, K(z,\bar z) + f(z) +
f^*(\bar z), \qquad q \kappa^2 \del_i K \to q \kappa^2 \del_i K + q
\del_i f \ee
In other words, although the Lagrangian is invariant under local
supersymmetry variation, it is not invariant under local K\"ahler
transformations. So in order to make the Lagrangian well-defined, we
have to cancel the K\"ahler variation of the covariant derivative by
accompanying it with a chiral $U(1)_R$ rotation on the fermions.

As a result, the superpotential, which has $R$-charge two under such
a rotation, is not a function on the moduli space, but rather a
section of a line bundle ${\cal L}_K$ on the moduli space, the line
bundle being defined by the transformation properties under K\"ahler
transformations above. The curvature of this line bundle is given by
the K\"ahler form \cite{Witten:1982hu}
\be c_1({\cal L}_K) = {i\over 2\pi} \del \delb K/{M_{Pl}^2} \ee
Moreover, this class should be quantized.

\newsubsection{Instanton correlators}
\subseclabel{InstantonCorrelators}

Suppose now instead we want to compute corrections to specific
correlators of chiral fields. To do this, we go back to the
partition function (\ref{PF}). It depends on a choice of vacuum,
which involves a choice of background ${ A}$ on $Z$. Let us consider
the partition function as a functional of the background field ${
A}$:
\be Z[{ A}]\ =\ \int dX^\perp\, d\Theta^\perp\, d\lambda\,
e^{-S[{A}]} \ee
In light-cone gauge, the couplings of the world-sheet fields to the
background gauge field and gaugino are given by
\be {\bf I} \ = \ \int_{}d^2 z \, {\rm Tr}({ A}_{\bar z} J_z) +
{1\over 4}( \Theta \Gamma_{\bar z}\Gamma^{m\bar n}\Theta) \,{\rm Tr}(
F_{m\bar n} J _z) +  j_{\bar z\widehat \alpha}{\rm Tr}(\xi^{\widehat \alpha}J_z) \ee
Here $\widehat \alpha$ refers to ten-dimensional spinor indices, and
$j_{\widehat \alpha} \sim \delb_{\bar z} X_M \Gamma^M
\Theta_{\widehat \alpha}$ is the supersymmetry current in light-cone
gauge. We will be interested mainly in isolated rational curves, in
which case the four-Fermi term vanishes.

The zero modes $\delta { A}$ are tangent vectors to the space of
gauge fields modulo gauge transformations at the chosen base point.
Since the unbroken supersymmetry generators relate bosonic and
fermionic wave-functions, we may collect them in $4d$ $N=1$
supermultiplets:
\be \delta {\bf A}_{\bar z}\ =\ (\phi  + \sqrt{2}\theta^\alpha
\chi_\alpha) \delta A_{\bar z} \ee
Thus the coupling of the worldsheet fields to the zero modes of
$(A,\xi)$ may be written as
\be {\bf I} \ \to \ \int d^2 z \, {\rm Tr}(\delta{\bf A}_{\bar z}
J_z) \ee
By differentiating, naively we find the $n$-point correlation
functions:
\be\label{Instcorr} \del_1 \ldots \del_n W\ =\ \int d\hat X\, d\hat
\Theta\, d\lambda\,\, {{\bf I}_1 \ldots {\bf I}_n}\, e^{-S[{\bf A}]}
\ee
By $W$ in this section we really mean $\Delta W$, the contribution
to the superpotential due to a worldsheet instanton wrapped on $\Q$.

Unfortunately these correlators are generally not well-defined on
Dolbeault cohomology classes, i.e. they depend on the representative
$\delta A_{\bar z}$. This does not mean that supersymmetry is
broken. The problem arises when insertions collide.\footnote{ This issue
was essentially previously encountered in \cite{Berglund:1995yu}:
because two-point functions were non-vanishing there, one could not
get a clean computation of certain Yukawa couplings. In the GS
computation one encounters the same issue: if one proceeds to
calculate higher point functions anyways, one finds the correlator
is not well-defined on Dolbeault cohomology classes and contact
terms are needed.}  To fix this, one
also has to add contact terms, which may be interpreted as Christoffel
symbols. This has the effect of covariantizing
the derivatives.

Geometrically, the chiral fields are sections of a
non-trivial bundle. As a result, an instanton correction to a
coupling of $n$ chiral fields appearing in the supergravity action
depends not only on the $n$-point functions, but also on all the
lower point functions and Christoffel symbols. In general one expects that if the superpotential
coupling is not forbidden due to too many fermion zero modes, it
will be non-zero. For instance if there are no left-moving zero
modes on the instanton, then the instanton generically contributes
to all possible couplings in the superpotential. This means that in
the generic situation, the instanton-generated superpotential itself
(the zero-point function) is the only natural quantity to compute.
This is of course exactly the computation done by
\cite{Buchbinder:2002ic}.

But there is still a way to check whether the instanton contribution
to a certain derivative of the superpotential is non-zero and we are
not missing any hidden cancellations. Namely we can try to tune the
bundle moduli so that we get extra left-moving fermionic zero modes.
For instance suppose we can tune the bundle moduli to get two
left-moving fermionic zero modes. For those values of the moduli,
the instanton then does not generate a contribution to $W$, and
\be \vev{D_i W}\ =\ \vev{\del_i W + \kappa^2\,\del_i K\, W}\ =\
\vev{\del_i W} \ee
so the $n=1$ case of (\ref{Instcorr}) can be interpreted as directly
computing the coefficient of a linear term $X_i$ in the
superpotential. By holomorphy, if this term is non-vanishing, then
it remains non-vanishing after a generic deformation. Similarly if
$\vev{W} = \vev{\del_k W} = 0$ then
\be \vev{D_2D_1W}\ =\ \vev{\del_2 D_1 W + \kappa^2 \, \del_2 K\, D_1
W + \Gamma_{21}^k D_k W}\ =\ \vev{\del_2 \del_1 W} \ee
and the $2$-point instanton correlator directly computes a mass term
in the superpotential. Again by holomorphy we conclude that if it
does not vanish for those special values of the moduli where we get
extra instanton zero modes, then it cannot vanish for generic bundle
moduli either.

A second situation where we can apply (\ref{Instcorr}) is when there
are `chiral' fermion zero modes on the instanton. All our fermions
were chiral anyways, but if $V$ is an $SU(n)$ bundle then some
left-movers are coupled to $V$ and others to $V^*$, so we can ask
for the net number of left-moving fermions valued in $V$ minus the
number of left-moving fermions valued in $V^*$. Equivalently, we can
ask for the net charge violation of the left-moving $U(1)$ symmetry.
As we already observed, this is governed by an index and vanishes
when $\Sigma_{37}$ is irreducible, so there is no net number of
chiral fermions in the above sense. However it may happen that
$\Sigma_{37}$ is reducible, with an equal and opposite number of
`chiral' zero modes on each irreducible piece. This can happen for
example if the vacuum admits an unbroken space-time $U(1)$ symmetry
and a gauged shift symmetry. In such a situation, again some of the
lower-point correlators are guaranteed to vanish, effectively
turning some of the covariant derivatives in (\ref{Instcorr}) into
ordinary derivatives. We'll discuss an example of this type in
section \ref{AnomalousU1}.

\newsubsection{Couplings in $E_6$ models}
\subseclabel{E6Couplings}

We would like to use (\ref{Instcorr}) to check for instanton
corrections to specific superpotential couplings. As a first
example, let us consider heterotic models with with an $SU(3)$
bundle, which yields $E_6$ GUT models in four dimensions. We
decompose the adjoint according to
\be {\bf 248}\ =\ ({\bf 1},{\bf 78}) + ({\bf 3},{\bf 27}) +
(\overline{\bf 3},\overline{\bf 27}) + ({\bf 8},{\bf 1}) \ee
Chiral fields live in $H^1(Z, {\rm adj}(E_8))$. Using the above decomposition,
one finds that the charged chiral fields are counted by
\be H^1(Z,V)\otimes {\bf 27}, \qquad
H^1(Z,\Lambda^2V)\otimes \overline{{\bf 27}\!} \ee
Let us see what kind of couplings we can compute.

We use the fermionic formulation of the $E_8\times E_8$ theory. In
this formulation, the 32 left-moving fermions transform are split
into two sets, each only transforming manifestly under an $SO(16)
\subset E_8$. Only bundles with holonomy contained in $SO(16) \times
SO(16)$ can be described in these variables. Embedding the $SU(3)$
holonomy group in $SO(16)$, the ${\bf 16}$ of $SO(16)$ splits as
\be\label{16decomp} {\bf 16} = ({\bf 3},{\bf 1})_{-1} + ({\bf \bar
3},{\bf 1})_{+1} + ({\bf 1},{\bf 10})_0 \ee
of $SU(3) \times SO(10)\times U(1)$. We will label the fermion
indices as $i,\bar i, a$ accordingly. In a $(2,2)$ model, the $U(1)$
may be identified with the left-moving $U(1)_R$ symmetry.

Under $SO(10) \times U(1) \subset E_6$, the ${\bf 27}$ decomposes as
\be {\bf 27} = {\bf 10}_{-1} + {\bf 16}_{1/2} + {\bf 1}_2 \ee
The vertex operators are of the form
\be\label{E6NSVertex} w(z)_I\wedge J^I = \, \delta A_{\bar
z,I}(z)d\bar z\wedge J^I \ee
For the ${\bf 27}$ states in the NS sector they are given by
\be\label{27VertexOp} V_{\bt}\ =\ w_{\bar z, i}(z)d\bar z\,
\lambda^{ i} \lambda^a, \qquad V_{\bf 1}\ =\ w_{\bar z,i}(z)d\bar z
\, g^{ i \bar j}\epsilon_{\bar j\bar k\bar l}\lambda^{\bar
k}\lambda^{\bar l} \ee
Let us also write the vertex operators for the bundle moduli, which are of the form
\be V_m\ = \ w(z)_{\bar z, I}d\bar z \, T^I_{i \bar j} \lambda^i
\lambda^{\bar j} \ee

We first consider generic $SU(3)$ bundles. Generically the
restriction of the bundle to $\Q$ is balanced, i.e.
\be V|_\Q \ = \  \cO(0) + \cO(0) + \cO(0) \ee
Recall that the left-moving instanton zero modes are counted by
\be H^0(\Q,{\cal V}(-1)|_\Q) \ee
where ${\cal V}$ is the rank 16 vector bundle transforming as
(\ref{16decomp}). Therefore in this case there are no left-moving
zero modes. As we already saw previously in section \ref{PfaffZero},
this implies that the instanton generically contributes to the
zero-point function (the partition function).  Therefore barring
unexpected cancellations, which would be hard to see and
non-generic, the instanton will contribute to all the couplings
appearing in the superpotential, but it would be hard to compute
them directly due to appearance of covariant derivatives rather than
ordinary derivatives. Let us now assume that we can tune the moduli
to get extra left-moving zero modes.

The next interesting case is
\be V|_\Q \ = \  \cO(1) + \cO(-1) + \cO(0) \ee
In this case we can reliably compute the one-point function. There
are two left-moving fermion zero modes, one for $\lambda^1$ and one
for $\lambda^{\bar 2}$. The wave functions for both of these zero
modes are constant on $\Q$. Vertex operators for vector bundle
moduli $X$ can absorb the zero modes if they are of the following
form when restricted to $\Q$:
\be
V_X \ = \ w_{\bar z, z}^X d\bar z\, \lambda^1 \lambda^{\bar 2}
\ee
This signifies the presence of a tadpole non-perturbatively:
\be
\del_X W \ = \ \int d\hat X\,d\hat \Theta\, d\lambda \, V_X\, e^{-S} \ \sim \left(\int_\Q w^X\right) e^{-T_\Q}
\ee
On the right we left out the one-loop determinants, which are
non-zero generically. If these tadpoles are non-vanishing, then
computations of non-perturbative corrections to higher order terms
in the superpotential can not be done cleanly in this perturbative
vacuum.

By a further tuning, we can get a splitting type of the form
\be\label{2-1-1Splitting} V|_\Q \ = \  \cO(2) + \cO(-1) + \cO(-1) \ee
This happens at complex codimension four in moduli space. In this
case there are four left-moving zero modes. An example of this is
the standard embedding, for which we have $V = TZ$. But in the
standard embedding we get the same splitting type above on any
generic isolated curve. For a more generic $SU(3)$ bundle satisfying
(\ref{2-1-1Splitting}), we would typically get this splitting type
only on $\Q$.

Zero and one-point functions are vanishing, and we can unambiguously
compute a two-point function with two vector bundle moduli (in other
words, non-perturbative mass terms for vector bundle moduli). But a
small surprise happens: the ${\bf 27}^3$ Yukawa couplings are
unpolluted by these non-vanishing two-point functions, so we can
also compute them. The reason for this is as follows. Let us denote
fields in the ${\bf 27}$ by $\Phi$ and neutral fields by $X$. We
look at the correlator
\be
D_\Phi^3 W \ = \ \del_\Phi^3 W + \kappa^2 \,\del_\Phi^2 K\, \del_\Phi W + \ldots +
\Gamma_{\Phi\Phi}^X  \del_\Phi \del_X W
\ee
The expansion in Christoffel symbols is quite messy, but the point
is that all these terms vanish if $\vev{W} = \vev{\del_X W} = 0$.
Terms like $\vev{\del_\Phi W}$ vanish automatically. As a result,
\be
\vev{D_\Phi^3 W } = \vev{\del_\Phi^3 W}
\ee
and so the ${\bf 27}^3$ Yukawa couplings can be computed even with
only four left-moving zero modes. We will do this computation next.
Note that this simplification does not hold for all Yukawa couplings
however. For the ${\bf 27}\cdot {\overline{\bf 27 \!}\,} \cdot {\bf
1}$ Yukawa couplings we do have to take account of the non-vanishing
two-point functions, as was previously found in
\cite{Berglund:1995yu}.

To make things slightly more transparent, let us use homogeneous
coordinates $\sigma^\zeta$, $\zeta = 1,2$ on $\Q$. We find the
following left-moving instanton zero modes,
\ba\label{LRFermionZero} \lambda^{ 1} &=& \sigma^\zeta
\alpha_\zeta \eol \lambda^{\bar 2} &=& \beta \eol \lambda^{\bar 3} &=& \gamma
 \ea
Recall that under $SO(10) \times U(1) \subset E_6$, the ${\bf 27}$
decomposes as
\be {\bf 27} = {\bf 10}_{-1} + {\bf 16}_{1/2} + {\bf 1}_2 \ee
so to get the ${\bf 27}^3$ we can try to compute the ${\bf 16} \cdot
{\bf 16}\cdot \bt$ or the $\bt \cdot \bt \cdot {\bf 1}$ coupling.
Both must give the same answer due to the underlying $E_6$ symmetry.
Let us first consider the $\bt \cdot \bt \cdot {\bf 1}$ coupling.

There are no bosonic zero modes or right-moving fermionic zero modes
other than the universal ones, so the zero mode measure is simply
given by
\be dM\ =\  d^2\alpha\, d\beta\, d\gamma \ee
Actually the true fermion measure also depends on the Pfaffian of
the non-zero modes, but again for our purpose this is not too
important. We have
\be d\sigma_i = \epsilon_{\alpha\beta}\sigma^\alpha_i
d\sigma_i^\beta = \vev{\sigma_i, d\sigma_i}, \qquad d^2\sigma =
d\sigma \wedge d\bar \sigma \ee
Using the vertex operators given above, we get
\ba
 \del_1 \del_2 \del_3 W &\propto&
 \int  dM\, \prod_{i=1}^3 d^2 \sigma_i \,
\vev{ V^{(1)}_{\bt}({\sigma_1})V^{(2)}_{\bt}({\sigma_2})V^{(3)}_{\bf 1}({\sigma_3})} \eol
 &\propto& \int  dM\, \prod_{i=1}^3 w^{(i)}(\sigma_i) d^2 \sigma_i \
 (\alpha \cdot  \sigma_1)(\alpha \cdot \sigma_2){1\over \vev{\sigma_1, \sigma_2}}\beta \gamma \eol
&=& \int \prod_{i=1}^3 w(\sigma_i) d^2 \sigma_i \ {\vev{\sigma_1,  \sigma_2}\over \vev{\sigma_1, \sigma_2}} \eol
 &=& \left(\int_{\Q} w^{(1)}\right)\left(\int_{\Q} w^{(2)}\right)
 \left(\int_{\Q} w^{(3)}\right) \ea
Here the propagator came from the non-zero modes
$\vev{\lambda^a(\sigma_1) \lambda^b(\sigma_2)} = {\rm
Tr}(T^aT^b)/\vev{\sigma_1,\sigma_2}$, and the fermionic integral
yielded $\int d^2 \alpha\,  (\alpha \cdot  \sigma_1)(\alpha \cdot
\sigma_2) = \vev{\sigma_1,\sigma_2}$.

Similarly we may try to compute the ${\bf 16} \cdot {\bf 16}\cdot
\bt$ coupling. The vertex operators for the ${\bf 16}$ create a
branch cut for the $\lambda$'s. We can eliminate the branch cut by
computing on the cylinder; equivalently we can work in a twisted
version of the theory. Let us first discuss some aspects of the R
sector states appearing in the ${\bf 248}$ of $E_8$; then by
breaking this down to various subgroups, we can get all others as
special cases.

Under $SO(16)$, the ${\bf 248}$ of $E_8$ splits as
\be {\bf 248} = {\bf 120} + {\bf 128} \ee
where ${\bf 128}$ is the positive chirality spinor of $SO(16)$. To
get the ${\bf 248}^3$, we compute the ${\bf 128}\cdot {\bf 128}\cdot
{\bf 120}$ coupling, which has two R sector states and one NS sector
state. Let us assume that the $SO(16)$ bundle $\V$ has a Hermitian
structure and can be split as $\V = W + W^*$. We consider the
following left-moving current:
\be J_L^{tot}\ =\ \sum_{i =1}^8 \lambda^{\bar i} \lambda^{ i} \ee
This corresponds to the `diagonal' $U(1)\subset SO(16)$. Now the
left-moving Ramond ground states form a Clifford algebra. The
`empty' state corresponds to the following vertex operator:
\be
\exp(\half \int J_L^{tot})
\ee
and the remaining states are found by acting on it with $\lambda^i$,
subject to a GSO projection. Hence we will twist by $\half
J_L^{tot}$. This maps the three-point function to a computation on
the sphere with only NS sector states. The new NS vertex operators
replacing the R sector vertex operators correspond to $(0,1)$ forms
valued in $\Lambda^{even} W$, and they naturally sit in the ${\bf
128}$.

After twisting, the left-moving fermions are sections of
\be\label{TwistedZeroCount} \Gamma(\Q, W|_\Q), \qquad \Gamma(\Q,
W^*\otimes K_\Q|_\Q) \ee
and the zero modes correspond to global holomorphic sections. From
Riemann-Roch, we expect at least eight left-moving zero modes, so
the calculation is going to be a bit different from the $\bt \cdot
\bt \cdot {\bf 1}$ coupling where we only had four zero modes.

In order to apply this to the case at hand, we further split
\be
W = V + U
\ee
where $V$ is our non-trivial $SU(3)$ bundle, and $U$ is  the trivial
rank five bundle. The fermions $\lambda^{ i}, i  = 1,2,3$ couple to
$V$, and the fermions $\lambda^{ b}, b=1,\ldots,5$ couple to $U$.
The vertex operators for the $\bt_{-1} \subset {\bf 27}$ where
listed in (\ref{E6NSVertex}). They correspond to $(0,1)$ forms
valued in $V \otimes (U + U^*)$. After twisting, the vertex
operators for the ${\bf 16}_{1/2} \subset {\bf 27}$ are given by
\be\label{E6RVertex}
\begin{array}{ccl}
  {\bf 16}_{1/2}: &  &  w(z)_{ i}\lambda^{ i}\lambda^b \\
   &  & w(z)_{ i
}\lambda^{ i}\lambda^{[b_1 b_2 b_3]} \\
   &  & w(z)_{ i}\lambda^{ i}\lambda^{[ b_1 b_2 b_3 b_4 b_5]}
\end{array}
 \ee
Here we used the shorthand notation $\lambda^{[b_1 ..b_k]} = \lambda^{b_1} \cdots \lambda^{b_k}$.
Also, after spectral flow the $U(1)$ charge shifts from ${\bf 16}_{1/2}$ to ${\bf 16}_{-1}$.
These vertex operators are made from $(0,1)$ forms valued in
\be
V \otimes (U + \Lambda^3 U + \Lambda^5 U)
\ee
Now when the bundle $V$ has splitting type $\{2,-1,-1\}$ on $\Q$,
from (\ref{TwistedZeroCount}) we find three zero modes for
$\lambda^{1}$, and one zero mode for each $\lambda^{ b_j}$, $j =
1,\dots,5$. Hence the correlator is simply
\ba
 \del_1\del_2\del_3 W &\propto&
 \int  dM\, \prod_{i=1}^3 d^2 \sigma_i \,
 V_{\bf 16}({\sigma_1})V_{\bt}({\sigma_2})V_{\bf 16}({\sigma_3}) \eol
 &=& \int  d^3\lambda^1_o \prod_{j=1}^5 d\lambda^{b_j} _o\prod_{i=1}^3 d^2 \sigma_i \,
 w^{(1)}_i(\sigma_1)\lambda^i \lambda^{b_1}\, w^{(2)}_j(\sigma_2)\lambda^j \lambda^{b_2}\,
 w^{(3)}_k(\sigma_3)\lambda^k \lambda^{[b_3b_4b_5]} + \ldots \ \eol
 &\propto & \left(\int_{\Q} w^{(1)}\right)\left(\int_{\Q} w^{(2)}\right)\left(\int_{\Q} w^{(3)}\right) \ea
The dots in the second line denote the other terms one gets by
writing out all the pieces of the vertex operators in the ${\bf 16}$
in (\ref{E6RVertex}) and in the $\bt$. To go from the second line to
the third line, first we absorbed one $\lambda^{1}$ zero mode with
each of the three vertex operators, so that the $w^{(i)}$ become
$(1,1)$ forms on $\Q$. For the $\lambda^{b_j}$ correlators, the only
terms that can be non-zero are the terms which have exactly one
$\lambda^{b_j}$ for every $j$. This follows from the unbroken gauge
symmetry, specifically the selection rules for the $U(1)^5 \subset
SO(10) \subset E_6$. Normally the correlator should be neutral under
each unbroken $U(1)$, but due to the twisting each $U(1)$ has a
background charge and so there is a shift by one. Finally then
absorbing the five $\lambda_o^{ b_j}$ zero modes picks the singlet
in the tensor product ${\bf 16} \otimes \bt \otimes {\bf 16}$. (In
terms of the decomposition above, this reduces to the familiar GUT
group algebra which yields the singlets $\bt \cdot \bt \cdot \bfb_h$
and $\bt \cdot \bfb_m \cdot \bfb_h$). So the correlator is
ultimately completely determined by the zero modes, and we get the
same answer as for the $\bt\cdot\bt\cdot{\bf 1}$ coupling, as it had
to be by the underlying $E_6$ symmetry.

We may also ask if we get a contribution to the $\overline{\bf
27}^3$. The vertex operators are similar to those in
(\ref{27VertexOp}) but the fermions are conjugated:
\be\label{27bVertexOp} V_{\bt} \sim w_{\bar z, \bar i}(z)d\bar z\,
\lambda^{\bar i} \lambda^a, \qquad V_{\bf 1} \sim w_{\bar z,\bar
i}(z)d\bar z\, g^{\bar  i  j}\epsilon_{ j k l}\lambda^{ k}\lambda^{
l} \ee
The fermion zero modes are still given by (\ref{LRFermionZero}). We
see that it is impossible to absorb the two $\lambda^{1}$ zero
modes. So the instanton contribution to the $\overline{\bf 27}^3$
vanishes. Non-vanishing instanton contributions can only come from
instantons with at most two left-moving fermion zero modes.

The calculations we have done here for the Yukawa couplings are
slightly simplified versions of calculations originally done  in
$(2,2)$ models in the world-sheet approach \cite{Dine:1987bq}. The
coefficient of the Yukawa coupling counts the number of holomorphic
maps from ${\bf P}^1$ to $\Q$ where three marked points are mapped
to $D_i \cap \Q$, with $D_i$ the Poincar\'e dual of $w^{(i)}$. As we
saw above, this gives the same number as the GS computation. The
main differences in the computation are as follows: we do not
integrate over the space of maps ${\bf P^1} \to Z$, only over the
space of (super)embeddings; and since we fixed the gauge redundancy
from the start, there are no (super)ghost correlators. As a result,
the GS computation is a bit shorter.

\newsubsection{Couplings in $SU(5)$ models}
\subseclabel{SU5Couplings}

By further breaking down $E_6$, we can generalize this to smaller
gauge groups.  We will briefly spell this out for $SU(5)_{GUT}$. In
the fermionic description, the manifest $SO(16)$ is broken to $SU(4)
\times U(1)$ by the $SU(5)$ holonomy, and the ${\bf 16}$ of $SO(16)$
is broken as
\be {\bf 16} = (\bfv,{\bf 1})_{+1} + (\bfb,{\bf 1})_{-1} + ({\bf 1},
{\bf 6})_0 \ee
under $SU(5) \times SU(4) \times U(1)$. Accordingly we split up the
16 indices as $i=1,\ldots,5$, $\bar i = 1,\ldots,5$, $a = 1,\ldots
,6$. Thus in this description, only a subgroup $SU(4)\times U(1)
\subset SU(5)_{GUT}$ of the GUT group is manifest. We use the
following decompositions of the matter representations under
$SU(4)\times U(1)$:
\be \bfv = {\bf 4}_{1/2} + {\bf 1}_{-2}, \qquad \bt = {\bf 4}_{-3/2}
+ {\bf 6}_1 \ee

Now let us think about the instanton zero modes. As usual they are
counted by
\be H^0(\Q,{\cal V}(-1)|_\Q) \ee
where in the present case
\be {\cal V} = V \oplus V^* \oplus \bigoplus_{a=1}^6 \cO \ee
The fermions
$\lambda^a$ are sections of $\cO(-1)_\Q$ and can never have any zero
modes. Let us denote by $V$ the $SU(5)$ bundle which breaks $E_8$ to
$SU(5)_{GUT}$. Then we can decompose
\be
V|_\Q \sim \bigoplus_{i=1}^5 \cO(a_i), \qquad \sum_i a_i = 0
\ee
Clearly we get essentially the same calculations as for the $E_6$
models, as the zero mode structure is completely determined by the
splitting type of the bundle on $\Q$. For instance let us try to
understand corrections to Yukawa couplings. To get a clean
computation, we need to arrange for four left-moving zero modes,
i.e.
\be a_i = \{2,-1,-1,0,0 \} \ee
Now we look at corrections to the $\bt\cdot\bt\cdot \bfv$ coupling,
which can be computed in the NS sector from the ${\bf 6}_1 \cdot
{\bf 6}_1 \cdot {\bf 1}_{-2}$. To find the vertex operators, we
recall the usual decomposition of the adjoint representation of
$E_8$ under $SU(5)_{H} \times SU(5)_{GUT}$:
\be {\bf 248} = ({\bf 24},{\bf 1}) + ({\bf 1},{\bf 24}) + (\bfv,\bt)
+ (\bfb,\btb) +(\bt,\bfb) + (\btb,\bfv) \ee
Therefore a state of the $\bfv$ matter representation comes from a
generator of the Dolbeault cohomology group $H^1(Z,\Lambda^2 V^*)$.
Similarly a state in the $\bt$ comes from a generator of
$H^1(Z,V)$. Hence the vertex operators are of the following form:
\be V_{\bf 1} = w(z)_{\bar z,\bar i\bar j}d\bar z\, \lambda^{\bar i} \lambda^{\bar j}, \qquad V_{\bf 6} =
w(z)_{\bar z, i}d\bar z\, \lambda^{ i} \lambda^a \ee
Essentially the same calculation as for the ${\bf 27}^3$ coupling above
shows that we get a non-zero contribution of the form $(\int_\Q w)^3$
of the instanton to the Yukawa coupling, since we have just broken
it into smaller pieces. We will not repeat it here.

Similarly we can try to compute the $\bt\cdot \bfb\cdot\bfb$
coupling for splitting type $ \{ 2,-1,-1,0,0 \}$, for instance by
evaluating the ${\bf 6}_1 \cdot \bar {\bf 4}_{-1/2} \cdot \bar{\bf
4}_{-1/2}$. Again due to the branch cuts it is convenient to compute
this in the twisted theory. This is similar to the ${\bf 16}\cdot
{\bf 16}\cdot {\bf 10}$ couplings in $E_6$ models. We again split
$\V$ as $\V = W + W^*$, and $W = V + U$ where $V$ is our non-trivial
$SU(5)$ bundle, and $U$ is a trivial rank three bundle. After
twisting, vertex operators in the Ramond sector become $(0,1)$ forms
valued in
\be \Lambda^{even}W = \bigoplus_{p+q=even} \Lambda^pV \otimes
\Lambda^q U \ee
I.e. we have vertex operators of the form
\be
\begin{array}{clccl}
  {\bf 4}_{-3/2}: & w(z)_{\bar z,i} d\bar z\,\lambda^{[ i b_1]} & \qquad &
  \overline{\bf 4\!}_{\,-1/2}: & w(z)_{\bar z,i  j}d\bar z\, \lambda^{[ i  j]} \\
   & w(z)_{\bar z,i }d\bar z\, \lambda^{[\bar i b_1 b_2 b_3]} &  &  & w(z)_{\bar z,i  j}d\bar z\,
\lambda^{[ i j b_1 b_2]}
\end{array}
\ee
and their conjugates. Here we have slightly abused notation,
indicating the $U(1)$ charge before spectral flow. After spectral
flow we have ${\bf 4}_{-3/2} \to {\bf 4}_{1}$, $\overline{\bf
4\!}_{-1/2}\to \overline{\bf 4}_{2}$. The calculation now proceeds
as before and is completely determined by the zero modes structure:
\ba
 \del^3W &\propto&
 \int  dM\, \prod_{i=1}^3 d^2 \sigma_i \,
 V_{\bf 4}({\sigma_1})V_{\bf 6}({\sigma_2})V_{\bf 4}({\sigma_3}) \eol
 &=& \int  \prod_{i=1}^3 d^2 \sigma_i \,
 w_{\bar z,z 4}(\sigma_1)w_{\bar z,z}(\sigma_2)w_{\bar z,z 5}(\sigma_3)\ \eol
 &=& \left(\int_{\Q} w_{ 4}\right)\left(\int_{\Q} w\right)\left(\int_{\Q} w_{ 5}\right) \ea

\newsubsection{Unbroken $U(1)$s}
\subseclabel{AnomalousU1}

The examples we have considered are somewhat uninteresting, in the
sense that the instanton generically has no left-moving zero modes,
and thus it should contributes to all the gauge invariant couplings
in the superpotential. The only really nice quantitity to compute in
this case is the Pfaffian. An exception to this rule appears in
vacua where extra left-moving zero modes are forced on us, such as
vacua with an unbroken $U(1)$ symmetry and a gauged shift symmetry.
The $U(1)$ symmetry does not necessarily need to be anomalous. We
will review some of the story here and state the $F$-theory
analogues in section \ref{M5Instanton}.

Let us assume that we have an unbroken $U(1)_X$ symmetry, and a
chiral field $X$ charged under it. If the symmetry is anomalous,
then the K\"ahler moduli will shift under a $U(1)_X$ gauge
transformation. The K\"ahler moduli and dilaton fields are defined
as
\be S\ =\  e^{-2\phi}V + i a, \qquad T_\Q\ =\
 \int_\Q J - {i}  \int_\Q B_2 \ee
Here $V = {\rm vol}(Z)$, $a=\int_Z B_6$, $\phi$ is the $10d$
dilaton, and volumes are measured in string units. Under a $U(1)_X$
gauge transformation we have
\be \delta S \ \sim \ i\,{{\rm Tr}(Q^X)\over 96 \pi^2}\,
\varepsilon_X \ee
where the trace runs over all the massless charged fields and $Q_X$
denotes the charge operator for $U(1)_X$, and
\be  \delta T_\Q \ \sim  \  \, i\varepsilon_X \int_\Q F^X \ =\ 2\pi
i\, q^X_\Q\, \varepsilon_X \ee
where we defined $F^X = {\rm Tr}( T^X \,F_{E_8})$. We will mostly
assume that ${\rm Tr}(Q^X) = 0$, although it is not hard to make
adjustments. If $\delta T_\Q \not = 0$, the $U(1)_X$ will pick up a
mass through the St\"uckelberg mechanism:
\be {\mathscr L}\ \ \supset \ \ (q_\Q^X A_{X\mu} -  \del_\mu {\rm
Im}(T_{\Q}))^2 \ee
This can be avoided by mixing with another $U(1)$ coming from the
second $E_8$.  At any rate, here we are mostly interested in
implications for instanton contributions to the superpotential.

If $q_\Q^X \not = 0$ then a worldsheet instanton wrapped on $\Q$
cannot contribute to the superpotential, because the classical
exponential factor is not gauge invariant:
\be e^{-T_\Q}\ \to\ e^{-T_\Q - 2\pi i q_\Q^X\, \varepsilon_X} \ee
Instead we will get a prefactor that is also not invariant and
cancels the gauge variation of the exponential. In this way, one
may generate superpotential terms which are forbidden at tree level
due to gauge invariance. But as long as the charged fields have
vanishing expectation value, this does not generate a potential for
$T_\Q$.

On the other hand, if $T_\Q$ shifts under a gauge transformation,
then we already get a potential perturbatively through the
$D$-terms. The K\"ahler potential is given by
\be {\cal K} \ = \  -M_{Pl}^2 \log (S + S^* - 2 q^X_S\,V_X)
-3\,M_{Pl}^2\, \log\, {1\over 6} \hat T_i \hat T_j \hat T_k
\,d^{ijk} \ee
with
\be \hat T_i\ =\ {\rm Re}\,T_i -  q^X_i\,V_X \ee
Assuming $q_S^X \sim {\rm Tr}(Q^X)=0$, from this K\"ahler potential
we get
\be \xi^X  = \left.{\del {\cal K} \over \del V_X}\right|_{V_X=0}\
\sim\
{M_{Pl}^2 \over \int J\wedge J \wedge J } \int c_1(L^X) \wedge J \wedge J\ %+\ {q_S^X M_{Pl}^2\over S + S^*}
  \ee
The $D$-term potential is given by
\be V_D = \half Re(S)^{-1}\left(\xi^X - \sum Q^X_\varphi
|\varphi|^2\right)^2 \ee
Without further information about the superpotential, we do not know
if the $\varphi$ can get a VEV. At zero VEVs the modulus $q_i^X
{\cal K}^{ij}{\rm Re}(T_{j})$ picks up a mass through the $D$-term
potential. This is expected by supersymmetry, because we already saw
that the imaginary part picked up a mass through the St\"uckelberg
mechanism. In this case the mass to be of order $g_s M_s$. If
$\varphi$ gets a large VEV, then the $U(1)_X$ gauge boson will eat
the $\varphi$ field instead of a K\"ahler modulus, and the K\"ahler
modulus may become massless again. The massive $U(1)_X$ will have a
mass of order the KK scale, and we have to re-expand around the
correct vacuum.

Although an instanton wrapped on $\Q$ does not contribute to the
superpotential, it could generate charged couplings
non-perturbatively. Let $X$ denote a chiral superfield with charge
$q_\Q^X$. One of the simplest couplings that could be generated this
way is the Polonyi superpotential
\be W\ =\ \mu^2 X e^{- T_\Q} \ee
Actually this is not an honest Polonyi superpotential but rather a
mixing between $X$ and $T_\Q$. At any rate, let us see how such a
term could get generated.

We first review how to construct an extra $U(1)_X$ symmetry. We
assume for simplicity that the holonomy is contained in $SU(n)$, and
take the heterotic bundle $V$ to be decomposible:
\be
V = V' + V''.
\ee
Since $c_1(V) = 0$, we have
\be
{\rm det}( V') = {\rm det}(V'')^{-1}
\ee
The extra $U(1)_X$ symmetry is the subgroup of $SU(n)\subset E_8$
that commutes with the holonomy of $V$. The internal part of the
gauge field of this $U(1)_X$ is a connection on a non-trivial line
bundle on $Z$, given by $L_X = \det(V')$.

Charged chiral fields correspond to generators of $H^1(Z, {\rm
ad}(E_8))$. It is not hard to see that the only fields $X$ that are
neutral under the GUT group but charged under $U(1)_X$ correspond
to a generators of $\Ext^1(V',V'')$ or $\Ext^1(V'',V')$. For
definiteness, let us say that $w^X \in \Ext^1(V'',V')$. The vertex
operator is of the form
\be
w^X = w^X(z)_{\bar z,I}d\bar z\, T^I_{i\bar j} \lambda^i \lambda^{\bar j}
\ee
where $i$ is an index valued in $V'$ and $j$ is an index valued in $V''$.

We assume that $\delta T_\Q \not = 0$, so that an instanton wrapped
on $\Q$ does not contribute to the superpotential. Now we want to
check for a coupling of the form $\Delta W = X e^{-T_\Q}$. In this
case, we are supposed to compute the correlator
\be \del_X W\  \propto\  \int dM \int_\Q  i^*w^X_I \wedge \vev{J^I} \ee
where $J^I$ is the current $\lambda^i \lambda^{\bar j}$, which
carries a non-zero $U(1)_X$ charge. Clearly this vanishes unless we
have precisely two fermionic left-moving zero modes to kill the
current, and absorb the $U(1)_X$ charge: one zero mode for
$\lambda^i$ for some $i$, and one for $\lambda^{\bar j}$ for some
$j$.

As usual, the bundle $V$ splits when restricted to $\Q$. Since
$\int_\Q F^X \not = 0$, the restriction $V|_\Q$ cannot be balanced.
We decompose
\be
V'|_\Q = \bigoplus \cO(a_i), \qquad  V''|_\Q = \bigoplus \cO(b_j), \qquad \sum a_i + \sum b_j = 0
\ee
In the minimal case, we have
\be \int_\Q c_1(V') = \int_\Q c_1(L^X) = 1, \qquad \int_\Q c_1(V'')
= -\int_\Q c_1(L^X) = -1 \ee
Generic bundles $V'$ and $V''$ with this property have splitting
types
\be
a_i = \{1,0, 0,... \}, \qquad b_j = \{ -1,0,0,..\}
\ee
for our rational curve $\Q$. This yields precisely the right number
of left-moving zero modes on $\Q$. We have essentially met such a
splitting type before in section \ref{E6Couplings}, however here the
presence of zero modes is guaranteed by $\int_\Q F^X \not = 0$ and
the splitting type is preserved under a generic deformation. If
$\int_\Q c_1(L^X) > 1$, there would be additional zero modes and we
would not be able to generate a term of the form $X e^{-T_\Q}$, but
we might be able to generate $X^m e^{-T_\Q}$ with $m>1$, a Yukawa
coupling, or a higher dimension operator like the Weinberg operator.

Suppose that both $V'$ and $V''$ are constructed from a spectral
cover. From the point of view of the spectral curve, $\Sigma_{37}$
has split into two pieces:
\be \Sigma_{37} \ = \ \Sigma_{37}' \cup \Sigma_{37}'', \ee
where
\be \Sigma_{37}' = \pi_Z^{-1}(\Q)\cap C', \quad \Sigma_{37}'' =
\pi_Z^{-1}(\Q)\cap C'' \ee
In the minimal case we get one zero mode on $\Sigma_{37}'$ with
charge $+1$, and one zero mode on $\Sigma_{37}''$ with charge $-1$.
A somewhat similar splitting phenomenon seems to occur when we take
the IIb limit in $F$-theory, as we will discuss later.

\newpage

\newsection{$M5$ instantons}
\seclabel{M5Instanton}

As in the heterotic string or type IIb, the superpotential in
$F$-theory satisfies a simple non-renormalization theorem. Let us
briefly recall the argument. Supersymmetry implies that the K\"ahler
moduli space is a complex manifold. Given a basis of four-cycles
$S_i$ in $B_3$, holomorphic coordinates on the moduli space are
given by
\be T_i\ =\ {\rm vol}(S_i) + i \int_{S_i} C_4 \ee
with volumes measured in ten-dimensional Planck units. Note that the
$T_i$ shift under gauge transformations of the RR field $C_4$, hence
such shifts are isometries of the K\"ahler moduli space. $F$-theory
is defined as a large volume expansion, so the small parameter in
$F$-theory is a K\"ahler modulus. For instance, the gauge coupling
constant $\alpha_{GUT}^{-1}$ is identified with the real part of
some $T_i$.

Now the superpotential must depend holomorphically on the $T_i$.
Then because of the shift symmetry, the superpotential can depend on
the $T_i$ only as
\be \exp (- T_i) \ee
In other words, the superpotential calculated by algebraic geometry
at tree level is actually exact to all orders in the expansion
parameter. It can receives corrections only non-perturbatively, for
example from $D3$-instantons. This is why $D3$-instantons are
important -- although small, they provide some of the leading
non-vanishing corrections to the superpotential.

We consider a $D3$ instanton wrapped on a four-cycle in $B_3$, the
internal manifold of our compactification. A precise way to define
an $F$-theory vacuum is to take an $M$-theory vacuum on a Calabi-Yau
four-fold $Y_4$ that is elliptically fibered over $B_3$, and then
take the limit in which the elliptic fiber shrinks to zero. In this
description, the $D3$ instanton in $B_3$ descends from an $M5$
instanton in $Y_4$ wrapping the $T^2$-fiber, and 7-branes are
encoded in the geometry of the elliptic fibration.

As mentioned in the introduction, the prescription of
\cite{Becker:1995kb,Harvey:1999as,Witten:1999eg} for computing the
leading instanton correction to the superpotential is to compute
the partition function obtained by integrating over the worldvolume
fields of the instanton. In this section we discuss how the
$M5$-brane sees the background geometry and fluxes. We then discuss
some general aspects of $M5$ instanton corrections in $F$-theory
compactifications. Although the leading contributions are in
principle computable, in practice this is a bit hard. Thus we
finally discuss instanton configurations where the computation of
the partition function and its derivatives can be reduced to those
discussed in section \ref{HeteroticInstanton}.

As a map for this section, it might be useful to keep the following
picture in mind. The partition function of the $M5$-brane is of the
schematic form
\be Z_{M5} \ = \ Z_\phi Z_\theta Z_{B^+} \ee
The behaviour of the partition function depends on the cohomology
groups of the $M5$-brane and their Hodge structure. The Hodge
diamond of an $M5$-brane is given by
\be
\begin{array}{ccccccc}
    &   &   & h^{3,3} &   &  &  \\
   &  & h^{3,2} &  & h^{2,3} & &   \\
   & h^{3,1} &  & h^{2,2} &  & h^{1,3} &  \\
 h^{3,0}  &  & h^{2,1} &  & h^{1,2} &  & h^{0,3}  \\
     & h^{2,0} &  & h^{1,1} &  & h^{0,2} &  \\
    & & h^{1,0} &  & h^{0,1} &  &  \\
    & &  & h^{0,0} &  &  &
\end{array}
\ee
The groups $h^{0,i}$ are related to the fermions $\theta$ on the
$M5$-brane, and $h^{0,3}$ is related to one of the scalars $\phi$.
The scalars and fermions are discussed in subsections 3.1 and 3.2.
The groups $h^{i,3-i}$ are related to the chiral two-form $B^+$.
This is the subject of subsection 3.3. Finally $h^{4}$ is related
to sources for the chiral two-form. This is discussed in section
3.4. Thus all the independent Hodge numbers play a role in the $M5$
partition function.

\newsubsection{Scalars and fermions}

The $M5$ worldvolume has $(0,4)$ supersymmetry and contains a single
superconformal tensor multiplet. Under the $SO(4)$ little group and
$USp(4)=SO(5)$ $R$-symmetry group, the supercharges transform as
$(1,2;4)$. The tensor multiplet consists of a self-dual two-form
$B^+$ transforming as  $(1,3;1)$, and five scalars $(1,1;5)$. The
fermions live in $(1,2;4)$ and satisfy a symplectic Majorana
condition (in Lorentzian signature). Reduction on $T^2$ yields the
$N=4$ Yang-Mills multiplet.

 The fermionic part of the
$M5$ action is of the schematic form
\be L_f^{M5} = \half \theta \left[ \not \! \nabla + \not \! G \right] \theta \ee
In order to understand the fermionic zero modes, we may first
consider setting $G$ to zero. The $G$-flux appears as a kind of mass
term, so the effect of turning $G$ back on will be to lift some of
the would-be fermionic zero modes.

Our $M5$ is wrapped on a divisor $D$ of $Y_4$. Accordingly, we split
the normal bundle as ${\bf R}^3 \oplus N$ where $N$ is a complex
line bundle. Since $N$ is the normal bundle to the divisor $D$ in
the Calabi-Yau four-fold, it follows from adjunction that $N$ is the
same as the canonical line bundle $K_D$ on $D$. Spinors on the $M5$
with chirality $(-1)^i$ are sections of
\be S^+_D \ = \ \bigoplus_{i\ {\rm even}}\, \Omega^{(0,i)}(D,{\bf
R}) \otimes K_D^{1/2} \ee
Since the tensor multiplet spinors have positive chirality on $D$,
we only take $i$ even. However as mentioned above, the tensor
multiplet spinors also transform as a ${\bf 4}$ under the $USp(4)_R$
symmetry group, i.e. the spinor of $SO(5)_R$. When the $M5$ wraps a
divisor $D$ in a Calabi-Yau four-fold, this is broken to $SO(3)_R
\times U(1)_R$ where the second factor is identified with the
structure group of the canonical bundle $K_D$ of $D$. The ${\bf 4}$
of $SO(5)_R$ splits as
\be {\bf 4}= {\bf 2}_{\half} \oplus {\bf 2}_{-\half} \ee
of $SO(3)_R \times U(1)_R$. Therefore, the tensor multiplet spinors
actually transform as
\be\label{M5Spinors} S^+_D \otimes (K_D^{1/2} \oplus K_D^{-1/2})\ =\
\Omega^{0,0}_{-\half} \oplus \Omega^{0,2}_{-\half} \oplus
\Omega^{3,0}_{+\half} \oplus \Omega^{3,2}_{+\half}
 \ee
under the $U(3)$ structure group of the metric restricted to the
divisor $D$, as well as the ${\bf 2}$ of $SO(3)_R$. The $\pm \half$
is to remind us of the $U(1)_R$ charges. The zero modes in the
absence of flux correspond to global holomorphic sections. Therefore
we get two universal fermionic zero modes from $h^{0,0}(M5)$, as
well as two non-universal zero modes for each generator of
$H^{0,i}(M5)$, $i=1,2,3$.

When we turn the $G$-flux back on, zero modes of Hodge type $(0,2)$
may get lifted \cite{Kallosh:2005gs,Saulina:2005ve}. The flux
induces a map
\be  \not \! G:\ \Omega^{0,2}(D) \to \Omega^{2,0}(D) \ee
Specifically
\be (\not \!G \theta)_{a  b}\ =\ \Omega_{abcn} G^{cn\bar e\bar f
}\theta_{\bar e \bar f} \ee
where $\Omega$ denotes the holomorphic $(4,0)$ form, restricted to
$D$. These equations can get further modified in the presence of
flux for the chiral two-form. The surviving fermionic zero modes are
in the kernel of this map. However we will mostly be interested in
situations where $H^{0,1}(M5) = H^{0,2}(M5) = 0$. In this case there
are no fermionic zero modes that could be lifted, and the presence
of $G$-flux is irrelevant in this regard.

Besides the fermionic zero modes, we also have bosonic zero modes.
The five scalars yield sections of $\Gamma(M5,N )\oplus {\bf R}^3$.
From this we recover three of the four universal bosonic collective
coordinates of the instanton, as well as non-universal bosonic zero
modes counted by $h^0(M5,N) = h^{0,3}(M5)$. The expectation value
of the chiral two-form $B^+$ through the $T^2$ decompactifies and
gives the remaining universal Euclidean ${\bf R}$ in the $F$-theory
limit. This zero mode is never lifted by the background $C_3$-field
as a three-form with zero or two indices in the $T^2$ does not
exist in $F$-theory. The remaining contributions of the chiral
two-form do not lead to vanishing of the partition function for
generic $C_3$, but they are much more complicated to understand.
They will be discussed separately in section \ref{ChiralTwoForm}.

\newsubsection{Fluxless backgrounds}

Since fluxes are small perturbations in the large volume limit we
may first find instantons and their zero modes for zero flux, and
then consider turning on the flux as a small perturbation.  $D3$
instantons in the absence of flux were studied by Witten
\cite{Witten:1996bn}. Let us review some of the results.

The $D3$ lifts to an $M5$ on a divisor $D$ in $Y_4$. The axionic
part of the K\"ahler modulus $T_D$ is not invariant under rotations
of the normal bundle. Indeed from the $11d$ supergravity action of
$C_3$
\be {1\over 2 (2\pi)^2} \int dC_3 \wedge * dC_3 + {1\over 2\pi}\int
C_3 \wedge I_8 + {1\over 6 (2\pi)^3} \int C_3 \wedge G \wedge G \ee
we find that
\be\label{C6Anom} \delta C_6\ =\ I_6^{(1)}(\Theta,R) + {1\over 4
\pi} G \wedge \Lambda \ee
so $T_D$ may shift under gauge and Lorentz transformations. Here
$\Lambda = 2\pi\varepsilon_X \omega^X$ where $\omega^X$ is a
generator of the coroot lattice, i.e. a generator in $H^2(Y_4)$
orthogonal to classes in $\sigma_{B_3 *}H^4(B_3)$ and
$\pi_Y^*H^6(B_3)$, and $\varepsilon_X$ is an infinitesimal gauge
transformation for the corresponding $U(1)_X$. It will be argued
that the shift in $\Theta$ is cancelled by an anomalous dependence
on $\Theta$ of the partition function of the fermions on the $M5$,
and the shift in $\Lambda$ is cancelled by the partition function
of the chiral two-form $B^+$.

In \cite{Witten:1996bn} it is conjectured that
\be\label{ExpThetaNVariation} \delta_{\Theta_N} \int_D C_6\ =\ -\Theta_N\, \chi(D,\cO_D) \ee
where $\Theta_N$ is an infinitesimal $U(1)$ rotation on $N$, the
normal bundle to $D$ in $Y_4$. Let us try to get this directly from
(\ref{C6Anom}). The anomaly polynomial is given by
\be I_8\  =\  -{1\over 48}\left[{ p_1(TD)^2 +p_1(ND)^2-2 p_1(TD)
p_1(ND)\over 4} -p_2(TD) -p_2(ND)\right] \ee
In our situation $ND = N \oplus {\bf R}^3$. Since we are interested
in a rotation of the normal bundle, we only need to look at the
pieces that depend on $N$. (The normal bundle will eventually be
identified with the anti-canonical bundle $K_D$, but we do not want
to do a simultaneous rotation on $TD$, as that would change the
$R$-charges of the fermions (\ref{M5Spinors}) and the chiral
two-form). Furthermore, $p_2(N)$ vanishes and $p_1(N) = c_1(N)^2$.
%Furthermore, for our background the holonomy takes values
%in $U(3)$, which we may identify with the structure group of $TD$,
%so we should be able to reexpress this in terms of the Chern classes
%of $TD$. By expressing everything in terms of the Chern roots,
%remembering that $N = K_D = \det TD^{-1}$, and recombining, we find
%that
%
%\be I_8\ =\ -{1\over 24} c_1(TD)c_3(TD) \ee
%
% Using $\half p_1 = -c_2 + \half c_1^2$, $p_2(N)=0$, the
%expression may be further simplified. Furthermore, by supersymmetry
%we should remember that $N = K_D = -\det TD$. Since we are
%interested in $\delta \int C_6$ under such normal bundle rotations,
Then we apply the descent procedure
\be I_8\ =\ dI_7^{(0)}, \qquad \delta_{\Theta_N}I_7^{(0)}\ =\ d I_6 \ee
The first step is ambiguous, because there is a one parameter family
of Chern-Simons forms with exterior derivative given by
$p_1(TD)p_1(N)$. We pick one on the criterion that $c_1(N)^3$ should
ultimately cancel, as it tends to give fractions with large
denominators. This leads to
% Depending on the form of $I_7^{(0)}$ we
%pick, we get
%%
%\be I_6^{(1)} \ \supset \  -\Theta_N \, {1\over 24}  c_2(TD) c_1(TD)
%\ee
%%
%
\be I_6^{(1)} \ \supset \  -\Theta_N \, {1\over 4\cdot
48}\left[c_1(N)^3 -2 \half p_1(TD) c_1(N) \right] =  +\Theta_N \,
{1\over  48} c_2(TD) c_1(TD) \ee
Modulo a factor of $-\half$, this recovers $-\Theta_N$ times the
index density for $\chi(D,\cO_D)$.
%However in
%general we get a linear combination of $c_1 c_2$ and $3 c_3$. It is
%not completely clear to us how to pick the right $I_7$. Since $\int
%c_3$ is the ordinary Euler character, requiring integrality might be
%a rationale for picking $c_2 c_1$.

The resulting anomaly of the exponential factor $\exp(-T_D)$ is
cancelled by the one-loop determinants of the worldvolume fields,
since the anomalies of the worldvolume fields cancel with the
anomaly inflow. Since the chiral two-form does not transform under
normal bundle rotations, the anomaly will come from the fermions.
From (\ref{M5Spinors}) we see that the net anomaly due to fermion
zero modes is given by $\chi(D,\cO_D)$, which cancels the anomaly
from the exponential factor precisely when
(\ref{ExpThetaNVariation}) holds. (By contrast, possible gauge
anomalies of $\exp(-T_D)$ due to (\ref{C6Anom}) are cancelled by
the partition function for the chiral two-form on the $M5$-brane,
as we will see later).

The four-dimensional supercharges carry charge $\pm \half$ under
such normal bundle rotations, so we identify this $U(1)_R$ with the
$4d$ $U(1)_R$ symmetry. In order to contribute to the
superpotential, we get the two universal fermionic zero modes which
each carry charge one-half, and generically no other fermionic zero
modes. Since the anomaly due to the universal zero modes $d^2\theta$
is equal to one, it follows that an $M5$-brane wrapped on $D$ can
only contribute to the superpotential if the remaining pre-factor has charge zero,
and therefore the exponential factor has
the opposite variation, i.e.
\be \chi(D, \cO_D)\ =\ 1 \ee

This analysis applies to smooth divisors. In $F$-theory we are
typically interested in cases where $D$ is not smooth. However when
we go to $M$-theory we expect that singularities of $Y_4$ can be
removed by a simultaneous resolution. We do not change the number
of zero modes in the $F$-theory limit, because it corresponds to
varying a K\"ahler modulus and the one-loop determinants we are
interested in don't have any dependence on the K\"ahler moduli.
Therefore let us perform such a resolution of $Y_4$ and analyse the
condition there.

Even when $Y_4$ is smooth, the divisor wrapped by the $M5$-brane
may sometimes have normal crossing singularities. The natural
prescription for such an $M5$-brane is to replace the cohomology
groups $H^k(M5)$ by the logarithmic cohomology groups
$H^k_{log}(M5)$ \cite{Donagi:2012ts,Clingher:2012rg}.

We would like to distinguish two cases:
\begin{description}
   \item[(i)]  The $D3$ worldvolume is not
contained in $\Delta$;
   \item[(ii)] The $D3$ worldvolume is contained in $\Delta$.
 \end{description}

Let us first consider type ${\it (i)}$ under heterotic/$F$-theory
duality. Such instantons get mapped to worldsheet instantons,
wrapped on a curve ${\cal Q}$ of genus $g_{\cal Q}$. In this case,
the $M5$-brane admits a fibration over ${\cal Q}$ whose fibers are
$K3$ surfaces. We can relate the Betti numbers and the Hodge
numbers of the $M5$ and ${\cal Q}$ through the Leray sequence.

We are mostly interested in instantons with only the two universal
fermionic zero modes, i.e. we want $h^{0,i}(M5) = 0$ for $i>0$. We
find that $h^{0,1}=g_{\cal Q}$ so to get $h^{0,1}=0$ we clearly
need $g_\Q=0$. If the fibration of the $M5$ over $\Q$ is
non-trivial then $h^{0,2}$ noramlly vanishes. To get $h^{0,3}=0$ it
is convenient to use the isomorphism $h^{0,3} = h^0(N)$ and check
if the normal bundle has sections, i.e. if the instanton is
isolated. Thus the arithmetic genus criterion for the $M5$-brane
agrees well with the heterotic picture. On the heterotic side, the
instantons that contribute are wrapped on a genus zero curve and
isolated, and using the arithmetic genus for the corresponding
$M5$-instanton we reach the same conclusions. Higher genus curves
would contribute to certain multi-fermion terms. Non-isolated
curves might possibly still contribute to the superpotential
(although probably they do not), but one has to understand how to
integrate over the family, just as for the $M5$-brane.

For more generic $M5$-instantons that only admit an elliptic
fibration it is harder to make universal statements. Generically we
will have $h^{0,1}=0$, and $h^{0,3}=0$ if the instanton is
isolated, but it is much harder to make a clear statement about
$h^{0,2}$ and it could well be non-zero. On the other hand,
$h^{0,2}$ might still get lifted by $G$-fluxes, so we might expect
fairly generic isolated $D3$ instantons wrapping a rational surface
to contribute to the superpotential. In anticipation of the next
section, we note again that even simple $M5$ branes intersecting
the discriminant locus typically have a large number of
three-cycles.

Next let us consider case ${\it (ii)}$. If the fiber type is of
type $I_1$, then the elliptic fiber has degenerated to a rational
curve. Since the curve has a double point, we should use the
logarithmic cohomology groups, which means that the arithmetic
genus criterion applies for our singular curve just as it applies
for a smooth curve.  The arithmetic genus of a nodal ${\bf P}^1$ is
the same as for an elliptic curve, i.e. equal to one. Therefore
$\chi(M5) \sim \chi(D3)\chi(T^2) = 0$. It follows that such an
instanton should not contribute to the superpotential in
$F$-theory.

This might seem somewhat counterintuitive from the IIb perspective,
as a $D3$-instanton on top of a 7-brane in IIb is generally thought
to contribute. The extra two fermionic zero modes implied by
$\chi=0$ are said to impose fermionic ADHM constraints, rather than
lead to vanishing of the superpotential contribution. It is
possible that we should be more careful because the $M5$ is
singular. However the prescription using the logarithmic cohomology
groups is natural as it behaves well under degenerations. Using
\cite{Clingher:2012rg} it matches with $D3$-instantons in the Sen
limit. Moreover if we further degenerate the fiber (which we
discuss next), we get results consistent with gauge theory
expectations.

Moving on, now let us assume that the fiber type is worse then
$I_1$, i.e. the $D3$-instanton wraps the same cycle as a 7-brane
with non-abelian gauge group $G$. In this case the $M5$-brane should
behave just like a gauge theory instanton. This can be verified from
the $M5$-picture, to some extent. The elliptic fiber over the $D3$
splits up into a chain of ${\bf P}^1$'s, one for each node of the
affine Dynkin diagram associated with the Kodaira fiber type,
satisfying
\be\label{AffineDynkin} \sum d_i \,[P_i] = [T^2] \ee
where $d_i$ are the Dynkin indices. If the singular locus is of
split type, then for each ${\bf P}^1$ we get a divisor $D_i$,
consisting of a ${\bf P}^1$-fibration over the $D3$ with fiber
$P_i$. For non-split type we get fewer such divisors, as some of the
$P_i$ are related globally by monodromy. The $M5$-brane can wrap
each of the $D_i$. In the context of $M$-theory compactified on such
a Calabi-Yau four-fold, these $M5$-branes may be identified with the
monopoles/fractional instantons of the non-abelian gauge theory.

In three dimensions the $N=2$ vector multiplet has an adjoint scalar
$\Phi$. Vacuum configurations satisfy $[\Phi,\Phi]=0$, so we may
diagonalize $\Phi$. Let us introduce ${\rm rank}(G)+1$ real scalars
denoted by $\phi_i$. Geometrically the $\phi_i$ specify the sizes of
$P_i$, which can be finite in the $M$-theory context, and they are
defined only up to Weyl transformations, which means we can restrict
them to take values in a fundamental domain. From
(\ref{AffineDynkin}) we have
\be\label{AffineConstraint} \sum d_i \phi_i\ =\ {\rm vol}(T^2)\
\equiv\ 1/R \ee
The ${\rm rank}(G)$ linear combinations of $\phi_i$ orthogonal to
this correspond to the eigenvalues of $\Phi$. Now let us assume that
the base is a del Pezzo surface, so that $h^{0,1}(S) = h^{2,0}(S) =
0$ and we get a pure $N=2$ gauge theory in three dimensions. The
superpotential of the three-dimensional gauge theory is given by the
partition function. We can again apply the Leray sequence, yielding
$h^{0,i}(D_j)=0$ for $i\not = 0$, so each $D_j$ contributes to
superpotential. Further if there are no monodromies among the $P_i$,
then $h^{3}(D_i)=0$ and the partition function for the chiral
two-form is trivial. Hence the partition function is given by
\cite{Katz:1996th}
\be W\ =\ \sum \exp(- d_i\phi_i/g_3^2) \ee
Since wrapped $M5$ branes correspond to monopoles, this is
reminiscent of the well-known results of Polyakov
\cite{Polyakov:1976fu}. Solving for the $F$-terms yields $h^\vee$
vacua, where $h^\vee$ is the dual Coxeter number of the non-abelian
gauge group $G$.

The relation of these vacua with the $4d$ gauge theory vacua arising
for $R\to \infty$ is somewhat subtle, but can formally be obtained
as follows. We introduce a Lagrange multiplier field $S$ to impose
the constraint (\ref{AffineConstraint}). The superpotential becomes
\be W \ = \ S(\tau - \sum d_i \phi_i/g_3^2) + \sum \exp(-
d_i\phi_i/g_3^2) \ee
where we used the relation $g_4^2 \sim R g_3^2$. Upon integrating
out the $\phi_i$ and using $\sum d_i = h^\vee - 1$ one obtains
\be W \ = \ \tau S + h^\vee\, S(\log (S/\Lambda^3) - 1) \ee
which is the Veneziano-Yankielowicz superpotential for the gaugino
bilinear $S \sim {\rm Tr}(\lambda\lambda)$. It is expressed purely
in terms of $4d$ quantities, so we can take the $3d \to 4d$ limit.
We can also integrate out $S$ to get
\be W \ = \ -h^\vee \Lambda^3 e^{-\tau/h^\vee}e^{2\pi i k/h^\vee}
\ee
in the $k$th vacuum. Although it looks constant, in field theory the
meaning of this superpotential is that $\Delta W$ calculates the
tension of domain walls. In a gravity theory, $\Lambda \sim M_{Pl}
\exp(-1/b_0 g_4^2(M_{Pl}))$ depends on the moduli and is not a
constant.

In the set-up above, there is another configuration we should
consider, namely an $M5$ wrapping the whole elliptic fiber. In fact
in the $F$-theory limit, we cannot wrap an $M5$-brane on each $D_i$
separately, but only on the sum $\sum d_i D_i$. Such an $M5$-brane
is singular, but the singularities are of normal crossing type, so
we use the arithmetic genus criterion. The holomorphic Euler
character of such a divisor vanishes, so the $M5$-brane does not
contribute to the superpotential. Indeed in the four dimensional
gauge theory this superpotential is not generated by instantons but
by strong dynamics, consistent with the arithmetic genus criterion
\cite{Witten:1996bn}. If the $M5$ had contributed, it would
indicate a stringy correction to the gauge theory result, which
should be absent (as it would modify the $E_8$ gaugino condensation
story for example).

Actually getting such pure gauge groups is somewhat rare. In
$M$-theory compactifications to three dimensions, we can certainly
construct ALE fibrations to get any desired gauge group. However in
order for such a compactification to lift to $F$-theory, the
compactification must admit an elliptic fibration, and this imposes
some important constraints. More typically, embedding in an elliptic
fibration will force the presence of matter curves, where the
elliptic fiber further degenerates. Such matter curves are closely
related to anomaly cancellation conditions in six dimensions. If the
matter curve has genus one or larger, and there is no $G$-flux, then
there are massive quarks which can becomes massless at special loci
on the moduli space.

Some special cases were studied in \cite{Diaconescu:1998ua}. Since
the $M5$-branes correspond to monopoles in three dimensions, one can
also understand their contributions more directly from the gauge
theory perspective \cite{deBoer:1997kr}. Taking the $R\to \infty$
limit should then yield the $4d$ superpotential.

\newsubsection{The chiral two-form and holomorphic factorization}

\subseclabel{ChiralTwoForm}

The most interesting field propagating on the $M5$-brane is the
chiral two-form. Let us first discuss its partition function from a
down-to-earth point of view, to see how theta-functions arise. Then
we move to a more abstract point of view. Most of the statements and
manipulations regarding theta-functions and holomorphic
factorization have well-known analogues on higher genus Riemann
surfaces. Mostly following
\cite{Witten:1999vg,Henningson:1999dm,Witten:1996hc}.

The basic problem with the theory of a chiral two-form is that it
can not have a conventional Lagrangian description. Indeed, suppose
we try to write one. By self-duality we have
\be \int_{M5} H \wedge * H \propto \int_{M5} H \wedge H = 0 \ee
We can however write the Lagrangian for the theory of a non-chiral
two-form, which contains both a chiral and an anti-chiral two-form,
in such a way that the anti-chiral part decouples from the chiral
part. Such an action is given by
\be\label{ChiralAction} S = \int_{M5} \half |H-i^*C_3|^2 - i H
\wedge i^*C_3 \ee
where $H$ is not required to be self-dual. The art is then to
compute observables using the non-chiral two-form, and to extract
the results for the chiral two-form by holomorphic factorization.
Let us discuss this procedure for the partition function.

In the partition function, we have to sum over the fluxes of $B$.
Let us choose a dual basis of 3-cycles $\{A_i,B^j\}$ in $H^3(M5,{\bf
Z})$:
\be
A^i \cap B_j = \delta^i_j
\ee
In a suitable basis, the $*$-operator has the following eigenvalues:
\be
*\omega^{3,0} = -i\, \omega^{3,0}, \quad
*\omega^{2,1} = +i\, \omega^{2,1}, \quad
*\omega^{1,2} = -i\, \omega^{1,2}, \quad
*\omega^{0,3} = +i\, \omega^{0,3}.
\ee
We take a basis $\omega_j$ of $ H^{2,1}(M5)+ H^{0,3}(M5) $, such that $*\omega_j = i\, \omega_j$.
Then up to an $SO(n)$ rotation we have
\be
\int_{A^i} \bar \omega_j = \delta^i_j, \qquad \int_{B_i}\bar \omega_j = \tau_{ij}
\ee
where $\tau_{ij} = \tau_{ji}$ is the period matrix.

Now we decompose $H = dB$ into a harmonic piece (the flux), which we
can think of as a classical field configuration,  and an orthogonal
piece which contains the quantum fluctuations. We may expand the
fluxes as
\be 2\pi\, n^i = \int_{A^i} H, \qquad 2\pi\, m_j = \int_{B_j} H \ee
%
%The fluxes may be either integer or half-integer; this depend on the choice of spin structure.
%We may write
%%
%\be
%H = 2\pi h^i \omega_i + 2\pi \bar h^i \bar \omega_i, \qquad C_3 = 2\pi z^i \omega_i + c.c
%\ee
%%
%where
%%
%\be
%h^i = -{i\over 2} (m_j-\bar \tau_{jk}n^k)(Im \tau^{-1})^{ij}, \qquad
%z^i = -{i\over 2} (x_j-\bar \tau_{jk}x^k)(Im \tau^{-1})^{ij}
%\ee
%%
We further write
\be
 C_3 = 2\pi\, z^i\, \omega_i + c.c
\ee

To obtain the classical partition function, we must evaluate the
path integral on all the classical field configurations:
\be Z_0(\tau,z)\ =\ \sum_{n_i,m^j} e^{-S_{cl}[\tau,z,m,n]} \ee
In the present context this sum was evaluated carried out in
\cite{Henningson:1999dm}. After Poisson resummation, and removing an
anomalous factor, it can be written as a sum of squares of theta
functions:
\be Z_0(\tau|z)\ \sim\ \sum_{\alpha,\beta} \left|\Theta \left[ {}^\alpha_\beta \right]
\!(\tau|z)\, \right|^2 \ee
where
\be \Theta [{}^\theta_\phi](\tau|z)\ =\ \sum_{{\bf Z} + \theta}
\exp(\half n^i n^j\, 2\pi i \tau_{ij} + 2\pi i n^i(z_i + \phi_i))
\ee
Since the action we wrote really describes both the chiral and
anti-chiral two-form, we now take a holomorphic square root to get a
partition function for the chiral two-form only. However there is no
unique choice for the classical partition function. Rather, there is
a unique partition function for every choice of spin structure.

In addition, we have to evaluate the path integral over the quantum
fluctuations and factorize the answer, formally described in \cite{Henningson:1999dm}. This
does not depend on the choice of spin structure. We get that the
partition function for spin structure $\alpha,\beta$ is of the form
\be Z^+[{}^\alpha_\beta](\tau|z)\ =\ {\Theta[{}^\alpha_\beta ](\tau|z) \over \Delta^+} \ee
where ${1/ \Delta^+}$
arises from the sum over quantum fluctuations. Since it is nowhere
vanishing and independent of the spin structure,
much of the interesting
information is contained in the classical theta-function.

We can also consider the more abstract point of view. The partition
function is a section of a line bundle on the intermediate Jacobian,
which we need to specify. We can do this the same way as for Riemann
surfaces. The Hodge $*$-operator induces a complex structure on the
Jacobian. This is also known as the Weil complex structure. The
intersection pairing induces a principal polarization $\omega$. The
line bundle is defined by specifying the phases $\phi =\pm 1$, subject to
\be \phi(a+b) = (-1)^{\omega(a,b)}\phi(a)\phi(b), \qquad
\phi(1) = (-1)^{2\alpha}, \quad \phi(\tau) = (-1)^{2\beta} \ee
which give the monodromies, and Chern class $\omega$, which is
positive definite. By the index theorem, for each spin structure we get
a unique theta function, corresponding to the unique
holomorphic section of the
line bundle, which we identify with
the partition function.

There is another well-known version of the Jacobian with a
different complex structure, known as the Griffiths Jacobian. It
corresponds to an involution with the same eigenvalue for
$H^{3,0}(M5)$ and $H^{2,1}(M5)$. The Griffiths Jacobian is known to
vary nicely in holomorphic families. For Calabi-Yau three-folds for
instance this can be seen by writing it as
\be {\cal J}(M5) \ = \ H^3_{\bf C}/F^2H^3_{\bf C} + H^3_{\bf Z} \ee
where $F^\bullet H^3_{\bf C}$ denotes the Hodge filtration on
$H^3(M5)$. It is a standard fact of Hodge theory that although the
individual $h^{p,q}$ do not vary holomorphically, the Hodge
filtration does, and thus by the above expression so does the
Griffiths Jacobian ${\cal J}(M5)$. However if $h^{3,0}$ and
$h^{2,1}$ are both non-zero, then $\omega$ is not positive
definite. As a result, instead of having a section, ${\cal L}$ has
higher cohomology and the theta function does not exist. If we
would try to construct it as a series as we did earlier but instead
with the indefinite norm, we would find that the series diverges.

The theta functions on the Weil Jacobian do not suffer from this, as
we might expect physically. However there is a price to pay; unlike
the Griffiths Jacobian, the Weil Jacobian generally does not vary
holomorphically in families.
%Its theta functions still satisfy the heat
%equation
%%
%\ba 0 & = & [{\del \over \del \tau_{ij}} + {\del^2 \over \del z_i
%\del z_j}] \theta \eol 0 & = & {\del \over \del \bar \tau}\theta \ea
%%
As we can see from the Calabi-Yau example above, in the Weil complex structure
the period matrix $\tau$ is not a holomorphic
function of the moduli. In other words, in general the partition
function of a chiral two-form suffers from a holomorphic anomaly.
The remaining contributions to the partition function from the
scalars and fermions appear to vary holomorphically.

Fortunately for isolated instantons, we have $h^{3,0}(M5)=0$ and
the Weil and Griffiths Jacobians coincide. So we do not get an
immediate contradiction with the holomorphy of the superpotential.
Non-isolated instantons might still contribute to the
superpotential after integration over the family. Perhaps we should
take this as evidence that non-isolated instantons in fact do not
contribute to the superpotential.

Finally, we should take an appropriate linear combination of these
partition functions together with those of the bosons and fermions,
so that we end up with a theory that is free from global anomalies
as well. The general prescription is not clear to us, but at least
in the cases we consider in detail there is a close relation with
the heterotic string and so there is a natural expression.

As an aside, the vanishing of the partition function of the chiral
two-form cannot be ascribed to its zero modes. The chiral two-form
is a bosonic field and has periodic identifications due to gauge
invariance, so it behaves as a compact scalar. Furthermore the
action does not depend on the VEV through a two-cycle. The integral
over these zero modes thus only gives a finite overall factor and
cannot cause vanishing of the partition function, so these zero
modes can be safely ignored. The one exception to this is the VEV
through the elliptic fiber, which becomes non-compact and is
identified with the emerging Euclidean ${\bf R}$ in the $F$-theory
limit.

\newsubsection{$U(1)$ symmetries and selection rules}
\subseclabel{M5GFlux}

All our above discussion assumed that $G|_{M5}=0$ in $H^4(M5)$. Is
this necessarily the case? Let us first recall that any $G$-flux in
an $F$-theory compactification must be orthogonal to four-cycles of
the following two types:
\begin{description}
  \item[(i)] $\sigma_{B_3 *}H^2(B_3) \subset H^4(Y_4)$
  \item[(ii)] $\pi^*_Y H^4(B_3) \subset H^4(Y_4)$
\end{description}
Now the class of the cycle wrapped by the $M5$ is itself in
$\pi^*_YH^2(B_3)$. Let us assume for now that $i_*i^*G =
\delta^2(M5)\wedge G$ is a non-trivial class in $H^6(Y_4)$. We will
return to the case that this fails later. Poincar\'e duality
implies that the intersection pairing $H^6(Y_4) \cap H^2(Y_4)$ is
non-degenerate. Then modulo torsion, under the above assumption we
have
\be G|_{M5}\not =0 \qquad \Leftrightarrow \qquad \delta^2(M5) \wedge
G \wedge \omega \not = 0 \ee
for some $\omega \in H^2(Y_4)$. Now $\omega$ cannot be in $\pi^*
H^2_Y(B_3)$ or $\sigma_{B_3 *} H^0(B_3)$, because then $\delta^2(M5)
\wedge \omega$ is a class of type {\it (i)} or {\it (ii)}, and the
$G$-flux would be automatically orthogonal to it. Any other $\omega
\in H^2(Y_4)$ is in the coroot lattice of the $4d$ gauge group, i.e.
$\omega = \omega^X $ for some $U(1)_X$ gauge symmetry. Therefore
\be G|_{M5}\not =0 \qquad \Leftrightarrow \qquad \int_{M5} G \wedge
\omega^X \not = 0 \ee
for some $\omega^X$ in the coroot lattice.

This leads us to the $F$-theory analogue of the gauged shift
symmetries that we encountered for the heterotic string in section
\ref{AnomalousU1} (also discussed in v2 of \cite{Donagi:2008kj}).
Recall from (\ref{C6Anom}) that a K\"ahler modulus $T_D$ shifts
under a $U(1)_X$ gauge transformation when
\be \delta T_D\ = \  {i\over 4\pi}\int_D G \wedge \Lambda\ =\ 2\pi
i\varepsilon_X \, {1\over 4\pi}\int_D G \wedge \omega^X \ = \ 2\pi
i q^X_D \varepsilon_X\ \not =\ 0 \ee
The low energy K\"ahler potential must then be of the form
\be {\cal K}_T \ \sim  \ -M_{Pl}^2\, \log {1\over 6\cdot 8}(T + T^*
- q^X V_X)^3 \ee
The $U(1)_X$ in this case picks up a mass through Green-Schwarz
couplings to the axion ${\rm Im}(T_D)=\int_D C_6$ and one finds an
Fayet-Iliopoulos term of the form
\be \xi^X\ \sim\ {M_{Pl}^2\over{\rm vol}(B_3)} \int_Y G \wedge
\omega^X \wedge
 J \ee
with volumes measured in $10d$ Planck units. Furthermore, a term in
the superpotential of the form
\be \int d^2\theta\, e^{-T_D} \ee
is forbidden by $U(1)_X$ gauge invariance. So an $M5$-brane wrapped
on $D$ can not contribute to the superpotential, but it can generate
couplings that are forbidden in perturbation theory due to the
$U(1)_X$ symmetry.

So let us consider an $M5$-brane with $G|_{M5}\not =0$ in $H^4(M5)$. In this case
there is a tadpole for the chiral two-form, because:
\be \int_{M5} H^+ \wedge C_3 \sim \int_{M5} B^+ \wedge G_4 \ee
and hence the partition function vanishes. We already saw the
meaning of this in the heterotic setting in section
\ref{AnomalousU1}; in this case there are chiral fermion zero modes,
and to get a non-vanishing answer we need to insert some fermionic
operators in the partition function to absorb these zero modes.

Here too there are some natural operators we have to insert to get a
non-vanishing answer. These are the Wilson surface observables
discussed in \cite{Henningson:1999dm}:
\be W(Q) = e^{i\int B^+ \wedge Q} \ee
where $Q$ is a four-form in $M5$. Dually we may think of $Q$ as a
two-cycle, and we will use the same notation to denote the
Poincar\'e dual. Since $\delta B^+ = \varepsilon_X \omega^X$, these
operators transform in the same way as charged chiral fields under
a gauge transformation:
\be W(Q) \ \to \ e^{i\varepsilon_X \int_Q \omega^X}\, W(Q) \ee
Furthermore if we now insert such operators in the path integral:
\be \vev{W(Q_1) \ldots W(Q_n)}\ \sim\  \int d[B] e^{i\int B^+
\wedge Q_1}\cdots e^{i\int B^+ \wedge Q_n} e^{ -S} \ee
and ensure that
\be \left[ {G\over 2\pi}\right]\ =\ [Q_1] + \ldots + [Q_n]   \ee
in $H^4(M5)$, then the tadpole for $B^+$ is cancelled and we get a
non-vanishing answer. Therefore these Wilson surface observables
play the role of charged fermionic zero modes for the $M5$-brane.
Again correlators involving such Wilson surface observables can be
computed (in principle) in the $M5$-brane theory using holomorphic
factorization.

In order to compute non-perturbative corrections to couplings of
charged chiral fields using the $M5$-instanton, we need the full
vertex operator corresponding to the chiral field. We may need to
dress up the operators $W(Q)$ above with extra factors for the
scalars and fermions. Even so it is not clear that explicit
calculations along these lines will be very illuminating, so we
will not proceed with this approach. Instead we focus on certain
special limits where the calculations can be be mapped to a more
standard problem.

Finally we return to the possibility that $\delta^2(M5) \wedge G =
0$ even though $i^*G$ is a non-trivial class in $H^4(M5)$. In terms
of Poincar\'e duals, this means that $[G|_{M5}] \in H_2(M5)$ is the
boundary of a three-chain $\Gamma$ in $Y_4$. By wrapping an $M2$
brane on $\Gamma$ with the opposite orientation, we can cancel the
tadpole for the chiral two-form. However even if this configuration
would be supersymmetric, it looks like a subleading effect, so at
order $\exp(-T)$ we do not expect it to contribute to the
superpotential.

\newsubsection{$M5$ wrapped on a four-cycle: the MSW CFT}

We have discussed a number of general aspects of $M5$-instantons,
but we saw that explicit computations, while possible in principle,
are not going to be easy. However the main goal of this paper is to
understand how $M5$-instantons interact with charged degrees of
freedom. For example we may have a Grand Unified gauge theory
localized on some divisor $S_{GUT}$ in $B_3$, and we want to
understand the non-perturbative corrections to it that may be
generated by an $M5$-instanton. For this purpose, it suffices to
take a certain weak coupling limit that was defined in
\cite{Donagi:2012ts}.

In the limit defined in \cite{Donagi:2012ts}, a local model
consisting of a $dP_9$ fibration over $S_{GUT}$ splits off from the
global model, while preserving the elliptic fibration over
$S_{GUT}$. The effective theory splits into a visible sector
involving the GUT, and a hidden sector associated to the rest of
the global model. In a perturbation expansion involving the
degenerating parameter $t$, these two sectors only talk through
non-charged interactions. For models with $K3$-fibrations, this is
the old heterotic $E_8\times E_8$ limit, but the limit of
\cite{Donagi:2012ts} applies equally well to models that do not
admit $K3$-fibrations.

The $M5$-instanton splits into two pieces in the limit,
\be M5 \ \to \  M5_{\rm vis}\cup_{\cal D} M5_{\rm hid} \ee
where ${\cal D} = M5_{\rm vis}\cap M5_{\rm hid}$. Now $M5_{\rm
vis}$ only talks to the charged degrees of freedom associated to
the Grand Unified model, and $M5_{\rm hid}$ only talks to the
charged degrees of freedom in the hidden sector. So we may
concentrate our efforts on $M5_{\rm vis}$. Furthermore, $M5_{\rm
vis}$ now admits a $dP_9$-fibration.

So we are now in the following situation: the $M5$ worldvolume is
fibered over a Riemann surface $\Q$:
\be
\begin{array}{ccc}
  S & \to & M5\\
   &  & \downarrow \\
   &  & \Q
\end{array}
\ee
The partition function depends only classically on the K\"ahler
moduli, through the exponential factor. As a result, we may change
the metric on the $M5$-brane without affecting the one-loop
determinants, as long as we keep the complex structure moduli
fixed. Thus we can scale down the fiber so that the $M5$-brane
collapses to a string, and consider the effective theory on this
string. This reduces the problem of computing zero modes and the
partition function to a problem on $\Q$, where we can address them
explicitly. Of course this is now starting to sound like the
discussion in section \ref{HeteroticInstanton}.

Let us do the reduction. The scalars simply reduce to non-chiral
bosons on $\Q$. As for the fermions, locally on $\Q$ we have $K_{M5}
\sim  K_S$ and
\be  \Omega^{0,2}(M5)\ \sim\ \Omega^{0,1}(\Q, \Omega^{0,1}(S))\
\oplus\ \Omega^{0,0}(\Q,\Omega^{0,2}(S)) \ee
Positive chirality spinors on the $M5$ are given by even $(0,i)$
forms. Upon reduction and a judicious use of the $*$-operator on
$S$, we get the following complex fermions
\ba
 & & S_\Q^+ \otimes \Omega^{0,0}(\Q,(\Omega^{0,0}(S) + \Omega^{0,2}(S)) \otimes {\bf 2}_{\pm \half} \eol
 & & S_\Q^- \otimes \Omega^{0,0}(\Q,\Omega^{0,1}(S) )\otimes {\bf 2}_{\pm \half}
\ea
When scaling down $S$, we should keep only the ground states on $S$,
i.e. the global holomorphic sections. Thus even degree Dolbeault
cohomology on $S$ yield right-moving complex fermions on $\Q$, and
odd degree Dolbeault cohomology yields left-moving fermions on $\Q$.
The symplectic Majorana condition reduces to an ordinary Majorana
condition.

Similarly we can understand the reduction of the chiral two-form. In
Euclidean space, the chiral two-form is imaginary self-dual.
Expanding in self-dual and anti-self-dual two-forms on $S$, and
taking ground states, we get $b_2^+(S)$ right-moving chiral bosons,
and $b_2^-(S)$ left-moving chiral bosons. Reduction with one or two
indices along $\Q$ yields only massive fields.

Finally we should also reduce the coupling to the $F$-theory
three-form $C_3$. Locally on $\Q$ this yields $b_2(S)$ gauge fields.
The coupling to the chiral two-form reduces as
\be \int_{M5} H^+ \wedge C_3 \ \to \ \int_\Q d^2 z \,\del_z \phi^+
A_{\bar z} \ee
The chiral bosons may be fermionized, and we end up with a
collection of chiral fermions coupled to gauge fields. This puts us
exactly in the type of situation studied in section
\ref{HeteroticInstanton}, where we know how to calculate an
instanton correction.

In particular, let us assume that $S$ is a $K3$ surface. In this
case we have $b_1(S) = 0$, $h^{0,2} = 1$ and $(b_2^-,b_2^+) =
(19,3)$. This is precisely the Narain data for the heterotic string
in $7$ dimensions, and together with the five non-chiral scalars and
the eight real fermions, we recover all the physical degrees of freedom
of the worldsheet theory of the $7d$ heterotic string.

In the context of $F$-theory, the $K3$ surface has some further
special properties. The elliptic fibration on the $K3$ surface
allows us to single out a $(1,1)$ sublattice, generated by the
elliptic fiber and a section:
\be v_1 = [T^2], \qquad v_2 = [T^2] + [{\bf P}^1] \ee
Of interest are the masses of membrane BPS states, wrapped on these
cycles.\footnote{In order to compare properly, we really have to
express the masses in $7d$ Planck units.} On the $M$-theory side,
these masses are proportional to the volume of the wrapped cycle in
Planck units. On the heterotic side, the $(1,1)$ sublattice
corresponds to the momentum and winding charges $(n_8,m_8)$ of a
distinguished $S^1$ on the internal $T^3$. Their masses are given by
\be m(v_1)\ =\ 2\pi n_8/R_8, \quad m(v_2)\ =\ m_8R_8/2\pi \ee
The BPS states with these charges are the Dabholkar-Harvey states.
We can follow the masses of these BPS states on both sides. In the
$F$-theory limit of small elliptic fiber with area $A \sim 1/R_8$,
the distinguished heterotic $S^1$ decompactifies, and the Narain
data reduces to that of the $8d$ heterotic string, i.e. a lattice of
signature $(18,2)$. Thus we may think of our MSW string as a
heterotic string living in the base of a $T^2$-fibration.

If $S=dP_9$ we still get a chiral string, but with a Narain lattice
of signature $(9,1)$. In the $F$-theory context, we further
eliminate the $(1,1)$ sublattice corresponding to the elliptic
fiber and a section of the $dP_9$. The resulting lattice of
signature $(8,0)$ is none other than the $E_8$ lattice. This can be
thought of as half of an eight-dimensional heterotic string, in the
limit that we decouple the winding modes on the internal $T^2$. We
could also consider an $M5$-brane wrapping $S = T^4$. In this case,
we get a left-right symmetric string.

Let us examine the case of a $dP_9$-fibration in more detail, since
this is what we encounter in the weak coupling limit of
\cite{Donagi:2012ts}. Then we can get much more insight by using
the spectral cover construction. Recall that the data of a $dP_9$
fibration over $\Q$ is equivalent to a spectral curve (previously
called $\Sigma_{37}$ in section \ref{HeteroticInstanton}) in a
complex surface ${\cal D}$ that is elliptically fibered over $\Q$,
where ${\cal D}$ is the log boundary of $M5_{\rm vis}$ above. There
is a direct mapping between an isotypic piece the cohomology of the
spectral curve $\Sigma_{37}$ and the `primitive' cohomology of the
$dP_9$ fibration, called the cylinder map \cite{Curio:1998bva}:
\be H^{p,q}_\Lambda(\Sigma_{37}) \to H_\Lambda^{p+1,q+1}(M5) \ee
Here $\Lambda$ refers to the eight-dimensional representation of
the $E_8$ Weyl group. (In general this breaks up further if the
structure group is less than $E_8$; this is why we encountered a
smaller piece in section \ref{HeteroticInstanton}). In particular
this maps ${\cal J}_\Lambda(\Sigma_{37}) \to {\cal J}_\Lambda(M5)$
isomorphically. We saw that the Prym actually played an important
role in the heterotic string, through the partition function of the
left-movers. This took the form of a theta function, and by the
above this corresponds to an analogous theta function for ${\cal
J}_\Lambda(M5)$ appearing in the partition function for the chiral
two-form $B^+$. Thus at least as far as the coupling to charged
degrees of freedom on $S_{GUT}$ is concerned, we have now
effectively reduced the computations for $M5$-instantons to those
for heterotic instantons in section \ref{HeteroticInstanton}.

We can similarly relate the story for the gauged $U(1)$ symmetries
for $M5$-instantons in section \ref{M5GFlux} to the analogous story
in section \ref{AnomalousU1}. A non-zero flux through $\Q$ gets
lifted to a non-zero flux though $\Sigma_{37}$, which gets mapped
to a $G_4$-flux through a four-cycle of the $M5$-brane using the
cylinder map. On the heterotic side we saw this corresponds to
gauging shift symmetries of the K\"ahler moduli and modifying the
zero mode structure of the worldsheet instanton. We saw an
analogous role of the intermediate Jacobian and the flux in the
$M5$-brane story in the sections 3.3 and 3.4. In particular with
this dictionary we can recognize the Wilson surface operators as
\be  e^{\int B^+ \wedge Q}=e^{\int_Q B^+} \ \to \ e^{\sqrt{2} Q
\cdot \phi^+} \ee
in the MSW worldsheet theory. Fermionizing $e^{\sqrt{2} Q \cdot
\phi^+}$ yields the charged fermionic zero modes of section
\ref{HeteroticInstanton}.

\newpage

\newsection{$D3$-instantons in the IIb weak coupling limit}
\seclabel{IIbLimit}

Let us recap what we have seen so far. $D3$-instantons in $F$-theory
are strongly coupled objects where both electric and magnetic
degrees of freedom are important, so we need some exact techniques
in order to deal with them. The situation becomes more transparent
when we interpret the $D3$ as a wrapped $M5$-instanton.

The worldvolume of the $M5$ contains a chiral two-form, which is
inherently quantum mechanical and does not admit a classical
description. Nevertheless there are some calculational techniques,
primarily the method of holomorphic factorization. We identified a
subclass of $M5$ instantons where we can further reduce calculations
to two dimensions, where we get a heterotic-like CFT description.
This connected our discussion on $M5$ instantons with our earlier
discussion on instantons in the heterotic string.

Most of the paper has been fairly conventional and unsurprising, at
least from the heterotic point of view. In this section we would
like to turn to the problem of understanding instanton corrections
from a more IIb like perspective, by taking a Sen limit. This
involves tuning the complex structure moduli and changing the
elliptic fibration at the location of the charged degrees of
freedom, and so we might expect some qualitative differences to
arise.

In fact there appears to be some tension between the behaviour of
$F$-theory instantons that we saw in this paper and $D3$ instantons
in perturbative IIb, especially in the $D3$-$D7$ sector. This can
be traced to the fact that the theory develops additional $U(1)$
symmetries as we dial the string coupling to zero. The extra
$U(1)$s impose strong selection rules in the IIb limit, but they do
not generalize to $F$-theory. This is already familiar for other
couplings, such as the classical $\bt\cdot\bt\cdot\bfv$ Yukawa
coupling in $SU(5)_{GUT}$ models, and we will see it holds for
instanton contributions as well.

Our results are therefore quite consistent with effective field
theory considerations: everything that is not protected by a
symmetry could in principle be generated. To illustrate this, we
show that some IIb selection rules on instantons in the presence of
chiral matter disappear in $F$-theory. This means that perturbative
type II doesn't capture some general properties of $M5$-instantons
that are relevant for phenomenological models. In addition the
issues of summing the contributions and multi-covers are not 
very much understood in the $D$-instanton approach. Thus it seems 
likely that further insights will have to come from a closer analysis
of heterotic instantons.

\newsubsection{General picture}

Let us briefly recall some aspects of open string quantization and
orientifolds in the `upstairs' picture, following the notation of
sections 13.4 and 14.3 of \cite{Polchinski:1998rr}. There is already
a discussion of instantons in IIb orientifolds available in the
literature, see \cite{Blumenhagen:2009qh} in particular, but we will
give a slightly extended discussion in this subsection and the next.
A closely related analysis has recently appeared in
\cite{Blumenhagen:2010ja}. We will use $\D3$ to denote instantons in
the upstairs picture and $D3$ to denote instantons in the downstairs
picture.

The $\D3$ branes live in a Calabi-Yau three-fold $X_3$, with a
holomorphic involution $\sigma$ mapping $\Omega^{3,0} \to
-\Omega^{3,0}$. The massless modes are of the form
\be \psi^M_{-1/2}\ket{0}_{NS}, \qquad \ket{s_0s_1s_2s_3s_4} , \qquad
\prod 2 s_i = -1\ee
In a curved background this leads to the following $33$ zero modes.
First we have the modes whose internal wave functions are given by
\be \Ext^0(i_*L, i_*L) \ = \ H^0({\cal D}3, \cO) \ee
Here we apply a $4d$ raising operator $\psi^\mu_{-1/2}$ to the
ground state, giving us the four real scalars $x^\mu$ describing the
position of the instanton. By supersymmetry, the same $\Ext^0$
generator defines a state $\ket{(-\half)^5}$ in the Ramond sector.
Together with $\ket{(+\half)^2(-\half)^3}$ this gives a fermionic
mode we call $\psi^{33}_{\alpha}$. From the complex conjugate of
this internal wave-function (or using the dual $\Ext^3$ generator)
we also get fermionic zero modes $\ket{(-\half )(+\half)^4}$ and
$\ket{(+\half)(-\half)(+\half)^3}$ which we call $\psi^{33}_{\dot
\alpha}$. Here $\alpha, \dot \alpha$ denote $SO(4) = SU(2) \times
SU(2)$ spinor indices associated to the uncompactified ${\bf R}^4$
directions.

By applying an internal oscillator $\psi^{\bar
i}_{-\half}\ket{0}_{NS}$, we further get $33$ zero modes whose
internal wave functions are given by
\be\label{33deformations} \Ext^1(i_*L, i_*L) \ = \ H^0({\cal D}3,K)
\oplus H^1({\cal D}3, \cO) \ee
These zero modes split up into modes that are even or odd under
$\sigma^*$.

The action of parity on the modes is given by
\be P:\alpha_m \to \pm e^{i\pi m} \alpha_m \ee
using $+$ for NN and $-$ for DD boundary conditions, and
\be\label{PsiParity} P: \psi_r \to \pm e^{i\pi r}\psi_r \ee
where the $\pm$ agrees with the action on $\alpha_m$ in the Ramond
sector, and is opposite in the NS sector. Thus the orientifold
action on the zero modes is almost the same as for the 77 zero modes
of a 7-brane wrapping the same four-cycle. The only essential
difference is that due to changing the boundary conditions from $DD$
to $NN$ along ${\bf R}^4$, from (\ref{PsiParity}) we get an extra
minus sign for some of the Ramond sector states. The action of
parity on the universal modes from $H^0(\D3)$ is then given by
\be x^i \to \gamma^{-1} x^{i\,T} \gamma, \qquad \psi_\alpha \to
\gamma^{-1} \psi_\alpha^T \gamma, \qquad \psi_{\dot \alpha} \to
-\gamma^{-1} \psi_{\dot \alpha}^T \gamma \ee
Hence for the $O(1)$ projection $\gamma = (1)$, we keep $(x^i,
\psi_{33}^\alpha)$ but project out $ \psi_{33}^{\dot \alpha}$.
Similarly, for generators of (\ref{33deformations}) we have
\be A \to - \gamma^{-1} A^{T} \gamma, \qquad \chi_\alpha \to
\gamma^{-1} \chi_\alpha^T \gamma, \qquad \chi_{\dot \alpha} \to
-\gamma^{-1} \chi_{\dot \alpha}^T \gamma \ee
For instance, to go from $\ket{(-\half)^5}$ to
$\ket{(-\half)(+\half)^2(-\half)^2}$ we apply two raising operators,
one with $NN$ and one with $DD$ boundary conditions. This relates
the parity action on $\psi_\alpha$ and $\chi_{\dot\alpha}$, with the
extra minus sign from (\ref{PsiParity}). One subtlety is that the
action on $\psi^{\bar i}_{r}$ is opposite for indices tangent to or
normal to the brane. However in (\ref{33deformations}) we have
mapped normal bundle valued forms to canonical bundle valued forms
by contracting with $\Omega^{3,0}$. Since $\sigma^* \Omega^{3,0}
=-\Omega^{3,0}$, the combined action of parity and $\sigma^*$ puts
all the generators of (\ref{33deformations}) on the same footing,
resulting in the orientifold actions listed above. Thus for even
generators of (\ref{33deformations}) we get fermionic zero modes
$\chi_{\alpha}$, and for odd generators we get bosonic and fermionic
zero modes $(A,\chi_{\dot\alpha})$, assuming the $O(1)$ projection.
Although listed for completeness, we are actually interested in
isolated instantons, so we assume there are no such zero modes.

Finally we need to quantize the $37$ strings. We will assume here that the $D3$ and $D7$
are intersecting, rather than coincident. In the upstairs picture, there are zero modes
with internal part given by
\be \Ext^1(i_*L_3,j_*L_7)\ \sim \  H^0(\Sigma_{37},L_3^\vee \otimes
L_7 \otimes K_{{\cal D}3}|_{\Sigma_{37}}) \ee
and similarly for the $73$ strings. This is similar to a $D1-D9$
system; in the $NS$ sector, the zero point energy is positive, so we
do not get any bosonic zero modes. In the Ramond sector the ground
state energy vanishes, but the raising operators carry positive
energy, so we cannot apply them to the ground state. Hence for each
$\Ext^1$ generator we only get a $2d$ chiral fermion $\lambda$ on
the intersection. The orientifold action relates the $37$ and $73$
zero modes.

We have to be a little careful about the bundle we put on the
$\D3$-instanton. We assume vanishing $B$-field in our discussion.
Recall that due to the Freed-Witten anomaly, the gauge field on a
3-brane does not take values in an ordinary line bundle but in a
`fake' line bundle
\be\label{FakeLine} \tilde L = L \otimes K^{-1/2} \ee
which is not necessarily integer quantized. Now world-sheet parity
relates a gauge field to its dual, and hence
\be P : \tilde L \ \to\  \tilde L^\vee = (L^\vee \otimes K) \otimes
K^{-1/2} \ee
We can also understand the extra factor of $K$ from the point of
view of $D$-brane charges. The coupling to RR fields is given by
\be \int_{X_3}{\bf ch}(i_*L)\,{\bf A}(X_3)^{1/2} \wedge {\bf C}
%, \qquad {\bf C} = \sum C_{RR}^{(i)}
\ee
where ${\bf C}$ is the formal sum of RR potentials. The extra factor
of $K$ yields the expected action on the Chern character,
viz.
\be {\rm ch}_j(i_*(L^\vee \otimes K))\ =\ (-1)^{j+1}\, {\rm
ch}_j(i_*L) \ee
This states, amongst other things, that $\D3$-brane charges ($j=3$)
are preserved. If we simply had replaced $L$ by $L^\vee$, this would
not have been the case.

Now assume we have an irreducible $O(1)$ instanton. The Euclidean
$\D3$-brane gets mapped to itself under the orientifold action, so
this implies that
\be \sigma^*(L^\vee \otimes K) = L \quad \Rightarrow \quad L\otimes
\sigma^*L = K \ee
Thus we can not take the trivial line bundle $\cO_{\D3}$ on the
$\D3$, but from (\ref{FakeLine}) we see that under favourable
circumstances it is compatible with `setting the gauge field to
zero.' Given a solution, we can obtain further solutions for any
generator of $h^{1,1}(\D3)_-$.

To summarize, the $D3$ partition function is schematically given by
\be
Z_{D3} \ = \ Z_\phi Z_\psi Z_F Z_{\lambda_{37}}
\ee
and we have discussed how the zero modes are related to various
cohomology groups. We would like to see if this structure is
reproduced from $F$-theory.

\newsubsection{Comparison}

Now let us start comparing this with an $M5$ instanton in the Sen
weak coupling limit. The coefficients of the Weierstrass equation
\be y^2 = x^3 + fx + g \ee
are written as
\ba f &=& -{1\over 48}( {\sf b}_2^2 - 24 \epsilon\,{\sf b}_4) \eol g
&=& -{1\over 864} (-{\sf b}_2^3 + 36 \epsilon\,{\sf b}_2 {\sf b}_4
-216\epsilon^2\, {\sf b}_6) \ea
with ${\sf b}_i$ a section of $K_{B_3}^{-i}$. Then we take a limit
$\epsilon \to 0$, so that the generic fiber becomes of type $I_1$.
In this limit, the elliptic fibration is given by
\be y^2\ =\ x^3 -{1\over 48}{\sf b}_2^2 -{1\over 864}{\sf b}_2^3\
=\ {1\over 864}({\sf b}_2 +6x) ({\sf b}_2 - 12 x)^2 \ee
Introducing a new coordinate $\tilde y = y/({\sf b}_2 + 12 x)$, we
can write this as
\be \tilde y^2 \ = \ {1\over 864}({\sf b}_2 +6x) \ee
This is the equation of a rational curve. The map $(x,\tilde y) \to
(x,y)$ identifies the two points
\be (x,\tilde y) \ = \ (-{\sf b}_2/12,\,\pm \sqrt{-{\sf b}_2/576})
\ee
on each fiber. Hence the generic elliptic fiber has degenerated to a
nodal curve, a ${\bf P}^1$ with two points identified. The two
points that get identified define a double cover $X_3$ over $B_3$,
branched over ${\sf b}_2 = 0$, which is a Calabi-Yau three-fold
because ${\sf b}_2$ is a section of $K_{B_3}^{-2}$. In other words,
we get a Calabi-Yau three-fold $X_3$ defined by an equation
\be \xi^2 = {\sf b}_2 \ee
and sitting naturally inside the Sen limit of the Calabi-Yau
four-fold. We identify $X_3$ with the IIb Calabi-Yau.

The intersection of our $M5$ with $X_3$ yields a divisor, which we
identify with the IIb $\D3$ instanton. This $\D3$-instanton is
similarly a branched double cover over a divisor $D3$ in $B_3$. It
is invariant under the orientifold action $\xi \to -\xi$, and
generically irreducible. Assuming it intersects the orientifold
locus, it should then correspond to an $O(1)$ instanton.

The relation between the $M5$-brane and the $D3$ it is fibered over
was largely discussed in section \ref{M5Instanton}, and we merely
need to take the limit. For the scalars, the normal bundle of the
$M5$ is the pull-back of the normal bundle of the $D3$, and they
have the same sections. There is one extra Euclidean normal
direction for the $D3$, which is identified with $\int_{T^2}B^+$
in the $M5$-description. In
particular the $D3$ inherits the four universal bosonic zero modes
from the $M5$. If the $D3$ is not tangential to the orientifold
locus, then the normal bundle to $\D3$ is the pull-back of the
normal bundle to $D3$. (Since the normal and canonical bundle are
the same upstairs, this also means that the difference between the
normal bundle and the canonical bundle to the $D3$ is given by the
intersection with the branch locus). This pull-back may have
additional sections, but the even sections (with respect to the
orientifold action) are the pull-back of the sections downstairs,
and recalling that the orientifold action on $H^0(\D3,N)$ and
$H^0(\D3,K)$ differ by a minus sign, from our earlier discussion we
see that these are precisely the ones that survive the orientifold
projection. The odd sections give rise to fermionic but not bosonic
zero modes after orientifold projection.

Next let us consider the fermions. We have essentially already
discussed this in section \ref{M5Instanton}, but let us also look at
this from the point of view of the Leray sequence for Dolbeault
cohomology
\be H^{i}(M5,\cO_{M5})\ \sim\ H^i(D3,\cO_{D3}) + H^{i-1}(D3,
R^1\pi_*\cO_{M5}) \ee
Here we simplified $R^0\pi_*\cO_{M5}\sim H^{0,0}(T^2)\otimes
\cO_{D3} = \cO_{D3}$. Close to the Sen limit, $R^1\pi_*\cO_{}\sim
H^{0,1}(T^2)\otimes \cO_{D3}$ is constant almost everywhere on the
base. Let us denote the generators of $H^1(T^2,{\bf Z})$ by
$\alpha,\beta$, subject to the relations $\alpha^2 = \beta^2 = 0$,
$\alpha \beta = -\beta\alpha =1$. Then the $(0,1)$ form on $T^2$ is
proportional to
\be \alpha_+\ =\ ({\rm Im}\tau)^{-1} (\beta - \bar \tau \alpha) \ee
The monodromy around an orientifold plane maps $\alpha_+ \to
-\alpha_+$, and the monodromy around a $D7$-brane leaves $\alpha_+$
invariant. We expect to see something similar in the Sen limit, but
the elliptic fiber has degenerated to a rational curve and there are
no ordinary one-forms left. Instead we have to deal with meromorphic
forms. Locally around the pinching $S^1$ we can write the equation
of the elliptic curve as
\be xy = \epsilon \ee
and the $(1,0)$ form as $dx/x$. In the limit $\epsilon \to 0$,
$dx/x$ becomes a meromorphic one-form. Resolving the double point,
we get a meromorphic one-form with opposite residue at the two
poles. Hence such meromorphic one-forms are related to functions on
$X_3$ odd under the involution.

Let us denote by $M5_\epsilon$ the $M5$ at finite $\epsilon$, $M5_0$
its Sen limit, and $\nu: \widetilde M5_0 \to M5_0$ the normalization
which separates the double point. Let us further define $\cO_{D3-}$
by
\be p_{D3*}\cO_{\D3} \ = \ \cO_{D3} \oplus \cO_{D3-} \ee
In other words, $\cO_{D3-}$ corresponds to functions on $\D3$ that
are odd under the orientifold involution exchanging the two sheets.
Then we have a short exact sequence:
\be 0 \ \to \ \cO_{M5_0} \ \to \ \nu_* \cO_{\widetilde M5} \ \to \
\cO_{D3-} \ \to \ 0 \ee
Applying the $R^i\pi_*$, i.e. taking cohomology along the fiber, we
get a long exact sequence which implies the isomorphism
\be R^1\pi_*\cO_{M5_0}\ \thickapprox\ \cO_{D3-} \ee
Therefore
\be H^{0,i}(M5)\ \sim\ H^{i}(D3,\cO_{D3}) + H^{i-1}(D3,\cO_{D3-})\
\sim\ H^{0,i}(\D3)_+ + H^{0,i-1}(\D3)_- \ee
in the Sen limit. In section \ref{M5Instanton} we saw that a
generator of $H^{0,i}(M5)$ gives rise to a fermionic zero mode
$\psi_\alpha$ for $i$ even and $\psi_{\dot\alpha}$ for $i$ odd. This
agrees precisely with the orientifold discussion. In particular, for
an isolated and rational $D3$ intersecting an orientifold plane, the
only expected non-universal fermionic zero modes come from
$h^{0,2}(M5)$.

Finally consider reduction of the chiral two-form. We label the two
one-cycles of the $T^2$ by $a$ and $b$. Reducing the chiral two-form
yields
\be\label{M5FormReduction} A^{elec} = \int_a B^+, \qquad A^{mag} = \int_b B^+ \ee
In the limit that the $a$-cycle pinches globally over the
worldvolume, we may think of the electric gauge field as elementary
and eliminate the magnetic gauge field. However in general neither
can be considered fundamental. The $Sl(2,{\bf Z})$ monodromies
generally kill the one-forms of the $T^2$ globally, so a single
$D3$-brane in $F$-theory usually does not carry any massless gauge
field. Equivalently in terms of the Leray sequence, $H^0R^1$ is
generically zero. This agrees with the IIb description, where a $D3$
intersecting the orientifold locus yields an $O(1)$-instanton. On
the other hand if the $D3$ does not intersect the discriminant
locus, $H^0R^1$ is non-zero and we get a $U(1)$ brane, as in type
IIb.

Assuming an isolated instanton $h^{0,3}(M5)=0$, the fluxes all live
in $h^{2,1}(M5) + h^{1,2}(M5)$. To understand the fluxes we need to
know more about $h^{1,2}(M5)$. Let us look at another Leray
sequence:
\be H^{1,i}(M5)\ =\ H^i(M5,\Omega^{1}) \ \sim \
H^i(D3,R^0\pi_*\Omega^1) + H^{i-1}(D3,R^1\pi_*\Omega^1) \ee
Here $R^0 \sim H^{1,0}(T^2)\otimes \cO_{D3} + H^{0,0}(T^2) \otimes
\Omega^1_{D3}$. Taking the Sen limit, the $(1,0)$ forms on $T^2$
become sections of the dualizing sheaf $\omega_P \equiv
\omega_{M5_0/D3}$, i.e meromorphic one-forms that may have poles
along the double point. Here we use $P$ to denote the fiber of
$M5_0$ over $D3$ which is a rational curve with double point, and
$\tilde P$ its normalization. We have a short exact sequence
\be 0 \ \to \ \nu_*\omega_{\tilde P} \ \to\ \omega_P \ \to \
\cO_{D3-} \ \to \ 0 \ee
The map from $\omega_P $ to $ \cO_{D3-} $ is the residue map. Taking
cohomology along the fiber, we get a long exact sequence that
implies
\be R^0\pi_*\omega_P \ \thickapprox\  \cO_{D3-} \ee
So we can write
\be H^i(D3,R^0\pi_*\Omega^1) \ \sim \ H^{0,i}(\D3)_- +
H^{1,i}(\D3)_+ \ee
For $R^1\pi_*\omega_P$ we will be very sketchy. The rough intuition
is that this is generally one-dimensional, but can jump up along the
locus where the $D7$-branes are located, because the elliptic fiber
should further degenerate there. The jump in dimension should
correspond to the number of coinciding $D7$-branes. Hence we would
get
\be H^{i-1}(D3,R^1\pi_*\Omega^1) \ \sim \ H^{1,i-1}(\D3)_- +
H^{0,i-1}(\D3)_+ + H^{0,i-1}(\D3\cap D7) \otimes n_{D7} \ee
We want to apply this for $i=2$. So what do all these pieces
correspond to? $h^{0,2}(\D3)_-$ corresponds to a $U(1)$ flux on
$\D3$ odd under the involution, but for supersymmetric
configurations this must vanish. $h^{1,1}(\D3)_-$ corresponds to an
odd $U(1)$ flux $F$ of type $(1,1)$, which is in principle allowed.
Fluxes proportional to $H^{1,2}_+$ and $H^{0,1}_+$ do not exist in
$F$-theory. Finally, a flux proportional to $H^{0,i-1}(\D3\cap D7)$
corresponds to a current $\bar J_\Sigma = \delb x\,
\delta^2(\Sigma)$, where $\Sigma$ is the intersection between the
$\D3$ and $D7$ branes.

In the IIb weak coupling limit, $C_3$ reduces to a 7-brane gauge
field $A_{D7}$ or the IIb two-forms $B_{RR}$ and $B_{NS}$. The $\int
C_3 \wedge dB$ coupling on the $M5$-brane reduces to couplings of
the form
\be \int_{M5} C_3 \wedge H \ \to \ \int_{\D3} A_{D7} \wedge
J_\Sigma\ + \ (B_{RR} + e^{-\phi} B_{NS})\wedge F_{\D3}  \ee
Thus the chiral two-form has the right couplings to describe the
$D3$-$D7$ strings in the IIb limit. By quantizing $D3$-$D7$ strings
using RNS we get $J_\Sigma$ as a fermion bilinear, as we saw
earlier.

Recall that the intermediate Jacobian was defined as $H^3(M5,{\bf
R})/H^3(M5,{\bf Z})$, and the partition function of the chiral
two-form was a theta-function on this Jacobian. In the IIb limit we
get extra structure: we get a projection to the anti-symmetric part
of the intermediate Jacobian of $\D3$, whose fibers are given by
the anti-symmetric part of the Jacobian of $\Sigma$ (i.e the Prym).
Thus if we fix the closed string data, then we recover the theta
function for the Prym of $\Sigma$, which is the partition function
for the chiral fermions on the $D3$-$D7$ intersection. Thus the
vanishing behaviour of the chiral two-form partition function is
usually controlled by the fermionic $D3$-$D7$ zero modes in the IIb
limit. If $h^{1,1}_-(\D3)\not = 0$ then it could also depend on the
expectation value of the $B_{RR}$ and $B_{NS}$.

Finally, we would like to briefly indicate how the results here may
be obtained more rigorously (and extended) using the approach of
\cite{Clingher:2012rg}. There we obtain a stable version of the
degeneration by blowing up Sen's family. The result is that apart
from the ${\bf P}^1$-fibration studied above (and called $M5_T$ in
\cite{Clingher:2012rg}), we get a second component $M5_E$ which is
a conic bundle over $D3$, and such that $M5_T \cap M5_E = {\cal
D}3$. Furthermore the cohomology groups of the smooth $M5$ become
the logarithmic cohomology groups of our degenerate $M5$-instanton.

Let us consider for example the Hodge numbers $H^{q,0}_{log}(M5)$,
which count fermionic zero modes. From the exact sequence
\be 0 \ \to \  H^0(\Omega^q_{M5}) \ \to \ H^0(\Omega^q_{M5}(\log
{\cal D}3)) \ \to \ H^0(\Omega^{q-1}_{{\cal D}3}) \ \to \ \ldots
\ee
we see that generators of $H^{q,0}_{log}(M5)$ may either descend to
$H^{q-1,0}_-({\cal D}3)$ by taking the residue, or lifted to
$H^{q,0}(M5)$. Using the Mayer-Vietoris sequence
\be 0 \ \to \ H^0(\Omega^q_{M5}) \ \to \ H^0(\Omega^q_{M5_T})
\oplus H^0(\Omega^q_{M5_E}) \ \to \ H^0(\Omega^q_{{\cal D}3}) \ \to
 \ \ldots\ee
and the ${\bf P}^1$-bundle structure, we further see that
$H^{q,0}(M5) = H^{q,0}(D3) = H^{q,0}_+({\cal D}3)$. So this
reproduces the relation between the fermionic zero modes of the
$M5$-instanton and the $D3$-instanton above. We can also see that
$H^{1,2}_{log}$ indeed reproduces the degrees of freedom of the
$D3$-$D7$ strings. This is discussed in more detail in
\cite{Clingher:2012rg}.

To summarize, we have examined the limiting form of the $M5$
partition function, which was given by
\be Z_{M5} \ = \ Z^M_{\phi}Z^M_\theta Z^M_{B^+} \ee
We saw that the $M5$ scalars reproduce five of the $D3$ scalars,
the $M5$ fermions reproduce the $D3$ fermions, and the chiral
two-form produces the $D3$ gauge field, the remaining scalar, and
the D3-D7 strings, i.e. schematically we have
\be Z^M_\phi\ \to\ Z_\phi^{(5)}, \qquad Z^M_\theta\ \to\ Z_\psi,
\qquad Z^M_{B^+}\ \to\ Z_\phi^{(1)}Z_F Z_{\lambda_{37}} \ee
This means that
\be Z_{M5} \ \to \ Z_\phi Z_\psi Z_F Z_{\lambda_{37}} \ee
and so modulo subtleties in properly defining the partition
functions, we reproduced the expected form of the $D3$ partition
function.

\newsubsection{Reducible instantons}

We would like to make some remarks about reducible $D3$-branes. The
situation here is very similar to degenerate 7-branes or spectral
covers. By Fourier-Mukai transform, it is also dual to bundles
constructed by extension.

In the previous section we saw that for an irreducible instanton,
the internal part of the zero modes is given by $\Ext^p(i_*L,i_*L)$
in the upstairs picture. Now suppose that we degenerate the $D3$ to
a reducible instanton, with the two reducible pieces intersecting
over the orientifold locus, either by varying instanton moduli or by
varying the ambient Calabi-Yau three-fold:
\be \D3\ =\ D3' \cup D3'' \ee
The sheaf on $\D3$ is represented by a pair of line bundles
\be L_3\ =\ (L_{3'},L_{3''})_{f_{glue}} \ee
supported on each component. Now generically, even though the
support is reducible, these two pieces are not independent, but are
glued by an isomorphism along the intersection $\Sigma_{3'3''} = D3'
\cap D3''$:
\be\label{fglue} f_{glue}\ \in\ \Hom(L_{3'}|_\Sigma,L_{3''}|_\Sigma)
\quad {\rm or} \quad \Hom(L_{3''}|_\Sigma,L_{3'}|_\Sigma) \ee
This gluing isomorphism may be interpreted as the expectation value
of a field localized on the intersection. The orientifold action
relates the two line bundles:
\be
\sigma^*(L_{3'}^\vee \otimes K_{3'})\ =\ L_{3''}
\ee
Since the orientifold action fixes $\Sigma_{3'3''}$, this means that
the gluing map is a section of
\be L_{3'}^{\vee}\otimes \sigma^*L_{3'}^{\vee} \otimes
\sigma^*K_{3'}|_{\Sigma_{3'3''}} \ee
or its inverse. We also map $f_{glue}$ to its dual, so it must
actually equal its dual and nowhere vanishing. Then we have
$L_{3'}^2|_{\Sigma_{3'3''}} = K_{3'}|_{\Sigma_{3'3''}}$. Furthermore
due to the orientifold symmetry we have $K_{3'}|_{\Sigma_{3'3''}} =
K_{3''}|_{\Sigma_{3'3''}}$. From the Calabi-Yau condition it then
follows that $K_{3'}|_{\Sigma_{3'3''}} = K_{\Sigma_{3'3''}}^{1/2}$
determines a spin structure on $\Sigma_{3'3''}$.

In this situation the $\Ext^0$,
which counts the universal instanton zero modes, is unchanged in the
degeneration limit. The reason for this is that although the support of the $D3$
becomes reducible, in the limiting configuration the naive zero
modes which we find on each irreducible piece separately must be
glued along the intersection $D3' \cap D3''$. As a result there is
only one $\psi_{\dot\alpha}$ in the upstairs picture, which is then
projected out by the orientifold action. Such a reducible instanton
has the same number of universal zero modes as an irreducible
instanton (i.e. four bosonic and two fermionic modes) and can
contribute equally well to the superpotential. We will call this a
reducible $O(1)$ instanton.

The number of $\Ext^1$ zero modes could jump, in conjunction with
$\Ext^2$ (because the Euler character does not jump). We get modes
from the $3'3'$ and $3''3''$ sectors, subject to the gluing
condition along the intersection. We also get modes in the
`off-diagonal' sectors. Since $\sigma^*$ fixes the intersection
$\Sigma_{3'3''}$, from our earlier discussion it follows that the
spinors on the $3'3''$ intersection live in
\be \Omega^p(\Sigma_{3'3''},L_{3'}^2 \otimes K_{3'}^{-1/2}\otimes
K_{3''}^{-1/2} \otimes K_\Sigma^{1/2})\ \sim \
\Omega^p(\Sigma_{3'3''},K_{3'} ) \ee
where we used the Calabi-Yau condition to get the second expression.
Therefore the internal wave functions of $3'3''$ instanton zero
modes (before projection) are counted by the global sections of
these sheaves. More generally we can use the $\Ext$-groups
\ba \Ext^1(\sigma^*P(i_*L_{3'}),\,i_*L_{3'}) &\sim&
H^0(\Sigma_{3'3''}, K_{3'}|_{\Sigma_{3'3''}}) \eol
 \Ext^1(i_*L_{3'},\,\sigma^*P(i_*L_{3'})) &\sim&
H^1(\Sigma_{3'3''}, K_{3'}|_{\Sigma_{3'3''}})^* \ea
These cohomology groups must have the same rank, as the Euler
character does not jump. Each generator gives rise to a certain
number of bosonic and fermionic zero modes. We get essentially the
field content of a hypermultiplet in a $D1-D5$ system. We have
\ba H^0(\Sigma_{3'3''}, K_{3'}|_{\Sigma_{3'3''}})& \to& X_{3'3''},\,
\xi_{3'3''}^{\dot \alpha},\, \xi_{3'3''}^{\alpha} \eol
 H^1(\Sigma_{3'3''},
K_{3'}|_{\Sigma_{3'3''}}) &\to &Y_{3'3''},\, \upsilon_{3'3''}^{\dot
\alpha}, \, \upsilon_{3'3''}^\alpha \ea
The orientifold action exchanges the $H^0$ and $H^1$ cohomology
groups, and thus kills precisely half the zero modes.

Let us consider again a reducible brane configuration but without
imposing an orientifold symmetry. It may be that the reducible
configuration has additional moduli for modifying the sheaf along
the intersection. In particular, let us consider the case where the
gluing morphism gets turned off. In this case, we end up with a pair
of line bundles supported on the two components of $\D3$, either
\be L_3 = (L_{3'}, L_{3''} \otimes \cO_{D3''}(-\Sigma_{3'3''}))
\qquad {\rm or}\qquad L_3 = (L_{3'}\otimes
\cO_{D3'}(-\Sigma_{3'3''}), L_{3''} ) \ee
depending on the direction of the gluing morphism. The sheaf $L_3$
is now rank two along the intersection. Notice that with $\tilde
L_{3''} = L_{3''} \otimes\cO_{D3''}(-\Sigma_{3'3''})$ we have
\be \Ext^1(i_{3'*}L_{3'}, i_{3''*}\tilde L_{3''}) %\ = \ H^0(\Sigma_{3'3''},
%L_{3'}^\vee \otimes \tilde L_{3''} \otimes K_{3'}|_{\Sigma_{3'3''}})
\ = \ H^0(\Sigma_{3'3''}, L_{3'}^\vee \otimes
L_{3''}|_{\Sigma_{3'3''}}) \ee
so comparing with (\ref{fglue}) we see that the gluing isomorphism
can be interpreted as the VEV of a field on the intersection, as we
said earlier. Conversely it shows that the fields localized on the
intersection do not correspond to deforming the support of the
branes.

In the limit of zero gluing map, one would get extra $\Ext^0$ and
$\Ext^1$ zero modes. Again this is familiar from reducible 7-branes
or degenerate spectral covers, where such a limit leads to enhanced
gauge symmetry and an extra chiral field. Let us denote the gauge
generators associated to each reducible branch by
$(\Lambda_{3'},\Lambda_{3''})$ and the chiral field on the
intersection corresponding to the zero mode $\delta f_{glue}$ by
$X$. The $\Ext^0$'s generate a relation among the $\Ext^1$'s:
\be\label{Ext0Relation} \delta X\ =\ \Lambda_{3'} X - X
\Lambda_{3''} \ee
When $X$ has a VEV, it gets eaten by the vector superfield
corresponding to $\Lambda_{3'} - \Lambda_{3''}$. In the limit of
zero VEV we get massless chiral and vector superfields obtained from
unhiggsing this massive vector superfield.

In the present context with $D3$ instantons, we get exactly the same
mathematical structure, but we interpret it as a (reducible) $U(1)$
instanton. However turning the Higgs VEV on or off looks like an
asymmetric operation which is not compatible with orientifolding.

When taking a Sen limit, the $M5$ typically limits to an irreducible
$D3$-instanton, but in special cases one may end up with a $D3$ with
reducible support. In particular it may happen that ${\sf
b}_2|_{D3}={\sf a}^2$ so that the $M5$ will factor into two pieces
$\xi = \pm {\sf a}$ in the Sen limit. It is not easy to follow the
gauge fields in the Sen limit, and it seems likely that generically
we get a reducible $O(1)$ instanton in the limit, rather than a
$U(1)$ instanton. If one does get a $U(1)$ instanton, this suggests
that $33$ zero modes on a reducible $U(1)$ instanton get lifted for
finite coupling and we end up with an $O(1)$ instanton in
$F$-theory.

\newsubsection{Comparison with heterotic}

Let us specifically consider the case of $SU(5)_{GUT}$ models. In
the neighbourhood of the GUT brane $S_{GUT}$, defined by the
equation $z=0$, the equations for the 7-branes become
\cite{Donagi:2009ra}:
\ba O7:\  0 & = & {\sf b}_2\ =\ b_5^4 + 4z b_4 \eol D7:\ 0 & = &
{\sf b}_8\  =\ z^5 (R(b_i) + P(b_i)z)  \ea
Here the $b_i$ are certain polynomials on $S_{GUT}$ which specify
the local model, and $R$ and $P$ in turn are certain explicitly
known polynomials in the $b_i$. The IIb Calabi-Yau three-fold
defined by $\xi^2 = {\sf b}_2$ has non-perturbative conifold
singularities at $z = b_5 = b_4 = 0$. A generic $D3$ instanton
intersecting the GUT brane will miss these singularities. We will
assume this is the case.

In the upstairs picture (before orientifolding), the locus $z=0$
consists of two pieces, $\xi = \pm b_5^2$, each wrapped by five
$D7$-branes. The intersection of the $D3$ with the GUT brane yields
a curve $\Q$ in the downstairs picture, and lifts to two copies in
the upstairs picture, which we denote by $\Q'$ and $\Q''$. In
heterotic/$F$-theory duals, the curve $\Q$ is the same curve on
which the heterotic worldsheet instanton is wrapped. The curve $\Q$
typically intersects the orientifold locus, i.e. it intersects the
matter curve $\Sigma_\bt$ on $S_{GUT}$.

In perturbative IIb, the GUT group is $U(5)$, and there are charged
$D3$-$D7$ modes transforming in the fundamental of $U(5)$ located on
$\Q'$. From the general discussion, the number of these zero modes
is computed by the Dolbeault cohomology group
\be\label{SU(5)37modes} H^p(\Q',L_3^{-1}\otimes L_7\otimes
K_{\D3}|_{\Q'}) \ee
where $p=1,2$, and $L_7$ is the $U(1)\subset U(5)$ line bundle on
the 7-brane containing $\Q'$. The orientifold action relates this to
similar cohomology groups on $\Q''$. We can further simplify this as
follows: if $\Q$ is an isolated rational curve, then its normal
bundle in $S_{GUT}$ is typically $\cO_\Q(-1)$. As long as $\Q$ is
not tangential to the branch locus, the normal bundle of $\D3$ is
simply the pull-back of the normal bundle to $\Q$ in $S_{GUT}$, so
we have $K_{\D3}|_{\Q'} = N_{\D3}|_{\Q'} =
\pi_{\D3/\Q}^*\cO_\Q(-1)|_{\Q'} = \cO_{\Q'}(-1)$. Now we consider
models where $c_1(L_7)$ very ample. This may lead to many generations
of $SU(5)$ multiplets, but that is besides the point here.
In such examples the $D3$-$D7$ zero modes on $\Q$ are guaranteed
to exist and are even completely chiral, because for sufficiently
ample $L_7$ the cohomology groups in (\ref{SU(5)37modes}) are
non-zero for $p=0$ and vanish for $p=1$.

Similarly the intersection of the $D3$-instanton with the flavour
D7-brane defined by $R + Pz = 0$ yields another curve that we will
call ${\cal R}$ in the downstairs picture. Its double cover is
generally irreducible, and we get chiral fermions living on ${\cal
R}$ with charge one under the $U(1)$ gauge group of the flavour
7-brane.

Now let us take $g_s$ finite. In this case we cannot use the Ganor
string approach, and the simplest description of the $D3$ instanton
is given by the $M5$-brane. We immediately see a number of
differences. First of all, the intermediate Jacobian of an $M5$-brane
in general can not be interpreted
as the Prym of any Riemann surface. Thus in general we have to use the
picture discussed in section \ref{M5Instanton}. In particular there
is no analogous formula for $D3$-$D7$
zero modes for finite coupling, there is not even a relevant
Riemann surface where such $D3$-$D7$ modes should live, and as we discussed
in the general formulation any
vanishing of the partition function is not due to zero modes of the
chiral two-form anyways. We should probably not have expected such
an analogous formula as in IIb, because there is no known higher dimensional
analogue of the Bose/Fermi correspondence in two dimensions.

The
closest relatives to (\ref{SU(5)37modes}) appeared in our discussion
on $dP_9$ fibered $M5$-branes, which appeared in the global-to-local limit.
As for the IIb limit, in that case the intermediate Jacobian of the $M5$-brane again degenerates and
we could reduce the problem to
two dimensions by using the spectral cover description of $dP_9$-fibrations.
In that description we encountered Dolbeault cohomology groups on
$\Q$ and $\Sigma_{37}$. But $\Sigma_{37}$ was different from $\Q'$
and these
cohomology groups were manifestly different from the
Dolbeault cohomology groups we encountered in the IIb limit.

Secondly, we found that these ``heterotic'' cohomology groups were
generically vanishing, even when the IIb cohomology groups can not
be. On $\Q$ we had $H^0(\Q,U(-1)|_\Q)$ where $U = \bigoplus_{i=1}^6
\cO(a_i)$ for $SU(5)$ models. For generic values of the bundle moduli
we further had $U = \bigoplus_{i=1}^6
\cO(0)$ (i.e. the bundle is balanced),
independent of discrete data like the number of generations.
But for such $U$
we do not get any zero modes; equivalently on $\Sigma_{37}$ we
found no zero modes generically. By holomorphy these facts imply that the
chiral two-form partition function is generically non-vanishing.
Nevertheless in the IIb limit it will vanish for the models considered above
due to the chiral $D3$-$D7$ zero modes we engineered. Of course this is not
a contradiction, since perturbative IIb models are basically special
boundaries on the moduli space of an $F$-theory compactification, and vanishing
on a special boundary does not imply vanishing elsewhere.

As the notation suggests, the curve $\Sigma_{37}$ (or rather its
image under the cylinder map) can be thought of to some extent as
the curve where ground states of $D3$-$D7$ strings are localized
for finite coupling, at least if we are dealing with $dP_9$-fibered
$M5$-branes.
In the spirit of \cite{Witten:1999eg}, suppose
we intuitively think of the worldsheet instanton as a
$D1$-instanton in type I, and the $10d$ gauge fields as living on
an $SO(16)$ $9$-brane. As long as we recall that our $SO(16)$ bundle
is further  embedded
in $E_8$, we can understand various aspects of the instanton this way.
Now we $T$-dualize along the $T^2$-fibers of
the heterotic Calabi-Yau three-fold $Z$. Then the $9$-branes are
mapped to $7$-branes wrapping the spectral cover, and the
$D1$-instanton is mapped to a $D3$-instanton wrapping
$\pi^{-1}_Z(\Q)$.

The spectral cover for the ${\bf 16}$ of $SO(16)$ has fewer pieces
than the spectral cover for $E_8$. For instance if we turn on
$SU(5)_H$ holonomy, we have the decomposition
\be {\bf 16} = (\bfv, {\bf 1})_{+1} + (\bfb, {\bf 1})_{-1} + ({\bf
1},{\bf 6})_0 \ee
under $SU(5)_H\times SU(4) \times U(1)$. The intersection of the
$D3$ instanton with the 7-branes is given by the union of $\Q$ with
multiplicity six, and two copies of $\Sigma_{37}$ (with different
spectral line bundles). When further embedded in $E_8$, the unbroken
gauge symmetry gets enhanced from $SU(4)\times U(1)$ to $SU(5)_{GUT}$,
and the spectral cover obtains additional components, which however
for our $SO(16)$ bundles store the same information as the pieces above.
We have already seen that integrating out
the physical modes on the instanton leads to the theta function of
$\Sigma_{37}$, as we would expect when integrating out the
fermionic $D3$-$D7$ strings.

\newsubsection{Zero modes in the IIb limit and $U(1)$ symmetries}

Consider again the IIB limit of one of our vertical
$D3$-instantons, which we assume to be irreducible in our
discussion, in the case of $SU(5)$ GUT models. In the last
subsection we saw that its contribution to the superpotential can
vanish in the presence of chiral matter, even if it contributes in
generic models with finite string coupling. This vanishing was seen
to be due to an extra $U(1)$ symmetry in perturbative IIb, which
promotes the $SU(5)$ gauge group to a $U(5)$ gauge group. Here we
would like to explore this behaviour a little more.

The extra $U(1)$ is generally lifted through its coupling to
axions. The mass obtained through this St\"uckelberg mechanism is
proportional to $g_s$, hence it is parametrically small compared to
the KK scale in the limit $g_s \to 0$, and so the $U(1)$ and its
selection rules are still visible in the effective action obtained from
KK reduction. However
as first pointed out in \cite{Donagi:2009ra,Pantev:2009de}, for
non-zero string coupling we cease to have a parametric separation
between the St\"uckelberg mass scale and the KK scale. So for
finite $g_s$ the extra $U(1)$ gauge boson should no longer be
distinguished, rather it should appear as one of the infinitely
many KK gauge bosons that one has in any case. (Note that this is a
generic argument and so may not always be true; we have seen that
one can have a St\"uckelberg mechanism also in $F$-theory).

Indeed in generic $F$-theory models there is no trace of this extra
$U(1)$ or its selection rules, as seen for example in the presence
of classical top quark Yukawa couplings, the computations of
\cite{Buchbinder:2002ic} or the example we will discuss at the end
of this subsection. But it is interesting to ask if we can also see
the $U(1)$ disappear explicitly under a small deformation away from
the perturbative IIb limit. If the $U(1)$ selection rules are to
disappear upon a small deformation, then there should be some
charged matter that screens the selection rules and lifts the mass
of the $U(1)_X$ gauge boson to the KK scale. It cannot be a
K\"ahler modulus, since the K\"ahler moduli in IIb survive in
$F$-theory. Still there are at least two ways this could happen.

The first is an $F$-theory version of section the mechanism in
section \ref{AnomalousU1}. We assume that the $M5$-instanton
generically contributes to the superpotential, in particular it has
no net $G$-flux on its worldvolume. In the $g_s \to 0$ limit, it
could happen that a four-cycle on the $M5$-worldvolume splits into
two four-cycles, with equal and opposite $G$-flux. Since we now
have net $G$-flux through topologically non-trivial cycles, we will
get vanishing behaviour. As in section \ref{AnomalousU1}, we expect
extra $U(1)_X$ charged matter in the limit.

The second mechanism involves certain non-perturbative conifold
singularities, such as those appearing in generic $SU(5)_{GUT}$
models from $E_6$ points when we take the Sen limit. As was seen in
\cite{Donagi:2009ra}, there is an extra charged half-hypermultiplet
localized at such a singularity which cannot be seen in string
perturbation theory. In some sense we should think of this matter
as arising from modes of the two-forms $B_{NS}$ and $B_{RR}$
through the vanishing ${\bf P}^1$ at this singularity. Indeed such
modes transform under the $U(1)_X$ gauge symmetry as
\be \delta \int_{{\bf P}^1}B\ \sim\ \varepsilon_X \int_{{\bf
P}^1}\omega^X \ee
A non-zero expectation value for the field $\exp \int_{{\bf P}^1}
B$ has the right form to generate classical top quark Yukawa
couplings in $SU(5)_{GUT}$ models when we turn on the string
coupling.

As in the previous mechanism, the $U(1)_X$ requires a longitudinal
mode from this charged matter and hence becomes a Kaluza-Klein
gauge boson in $F$-theory \cite{Donagi:2009ra}. The fields $B_{NS}$
and $B_{RR}$ come from the three-form $C_3$ in $F$-theory, so this
means that the charged fields above should come from expanding
$C_3$ along a differential three-form $\gamma$ such that $d\omega^X
= \gamma$ for finite $g_s$ and $\gamma$ is localized at the
conifold singularities in the limit $g_s \to 0$. Of course there
are infinitely many KK gauge bosons in $F$-theory and in general
there is no parametric separation between their masses, so unless
we want to discuss all of them we should integrate them out and
work with the effective action below the KK scale. The instantons
of this effective Lagrangian are no longer required to satisfy the
$U(1)_X$ selection rules. Alternatively we keep the KK modes. The
gauge variation of $C_6$ is still cancelled by the gauge variation
of $Z_{B^+}$, but since $Z_{B^+}$ now depends on charged fields
with a non-zero VEV, this does not imply the vanishing of
$Z_{B^+}$.

Let us ponder the implications of this observation for KKLT like
moduli-stabilization scenarios.\footnote{There are various
conceptual issues with these scenarios, most notably the use
perturbation theory is inconsistent if the number of vacua is indeed
finite non-perturbatively. We ignore these issues here and only
concentrate on perturbative vacua, where we can take the volume
modulus arbitrarily large.} It has been argued in
\cite{Blumenhagen:2007sm} that there is an inherent conflict between
the presence of chiral matter and instanton contributions to the
superpotential. The argument can be phrased as follows: gauge groups
in type IIb are either $U(n)$, $O(n)$ or $USp(n)$. Chiral matter is
associated with $U(n)$ groups. In order to get chiral matter, a flux
has to be turned on for $U(1)\subset U(n)$. Let us consider the
effect of a gauge transformation for this $U(1)$. The RR four-form
transforms as
\be
\delta_\lambda C_4\ \sim\ {\rm Tr}(\lambda F) \delta^2(S)
\ee
where $S$ is the four-cycle on which the $U(1)$ is localized. By
Poincar\'e duality there is some class $D$ in $H^2(X_3)$ such
that\footnote{This does not follow from what we said so far, and
could lead to counterexamples. But ${\rm Tr}(F)/2\pi$ should be in
the image $H^2(X_3) \to H^2(S)$, because we want non-zero
intersection with the matter curves.}
\be
\int_{S \cap D} {\rm Tr}(F) \not = 0
\ee
Under a $U(1)$ gauge variation the associated K\"ahler modulus transforms as
\be
\delta_\lambda\,  {\rm Im}(T_D) \ = \ \delta_\lambda \int_D C_4\ =\
\lambda \int_{S\cap D} {\rm Tr}(F) \not = 0
\ee
Therefore a contribution of the form
\be \int d^2 \theta\, e^{-T_D} \ee
is forbidden by gauge invariance. This is the behaviour we saw in
the IIb limit for $SU(5)$ GUTs. It may not be a problem, since in
this case the K\"ahler modulus is in fact eaten by the $U(1)$ vector
multiplet, but it certainly affects the potential for the K\"ahler
moduli.

As discussed above, for finite $g_s$ this extra $U(1)$ disappears
into the KK tower and there are charged fields with non-zero VEV,
so there is no $U(1)$ selection rule that could forbid a non-zero
contribution of the form $\int d^2\theta\, Z_{B^+}\, e^{-T_D}$ to
the superpotential. But one may wonder if we missed some effect and
there is still some a priori conflict with chirality.

Since there is no such conflict in the heterotic string,
heterotic/$F$-theory duality predicts that there can be no such
issue in $F$-theory either. Let us examine this more closely. The
K\"ahler moduli $T_\Q \sim {\rm vol}(\Q \times {\bf P}^1)$ on the
$F$-theory side can all appear in the superpotential. The last
modulus $T_{0} \sim {\rm vol}(\sigma_{B_2})$ looks more problematic;
wrapping a $D3$ on the GUT cycle yields a field theoretic instanton
and field theory index results seem to indicate zero modes if there
is chiral matter. However as we discussed earlier in section
\ref{ChiralTwoForm}, unless there are gauged shift symmetries,
$G|_{M5}=0$ even in models with chiral matter, so even $T_0$ should
appear in the superpotential.

On the heterotic side, a $D3$ instanton wrapped on $\sigma_{B_2}$
corresponds to a space-time instanton, and the K\"ahler modulus
turns into the $8d$ string coupling. So it looks like we would have
trouble generating an $\exp(-S)$ contribution to the superpotential,
where $S$ denotes the heterotic dilaton. However we could also
consider instantons in the second $E_8$, where gaugino condensation
leads to an $\exp(-S)$ term in the superpotential. (Strictly
speaking this contribution is not generated by instantons, but that
is not essential here). So there cannot be any a priori conflict
with chirality.

We can check this also directly on the $F$-theory side. The
$F$-theory analogue of this is as follows: we may also wrap a $D3$
at the infinity section of $B_3$, instead of the zero section. The
K\"ahler modulus for this cycle is $T_\infty \sim T_{0} + n_i
T_{\Q_i}$. The $G$-flux through this cycle at infinity is zero.
There is no need for chiral matter at this location (as we also know
from the heterotic string, because it corresponds to the second
$E_8$), and so from this $D3$-instanton we can still get a term
$\exp(-T_{0})$ in the superpotential.

\newsection{Conclusions}

In this section we would like to summarize some of the main lessons we have learned about
instanton corrections.

\begin{enumerate}
  \item In order to compute instanton corrections, we first need to settle
  on a reasonable prescription. For instantons of the type considered here we know of only
  one prescription, which was pioneered in \cite{Becker:1995kb,Harvey:1999as,Witten:1999eg}
  and is usually called the $D$-instanton approach
  or the physical gauge approach. It asserts that up to factoring out some universal zero modes,
  the contribution
  of an instanton to the superpotential is given by its partition function
  $Z$, obtained by integrating over all the
  physical fluctuations.

\item For an $M5$-brane instanton the partition function
  is of the form $Z_{M5} = Z_{\phi}Z_\theta Z_{B^+}$. Of the three factors in $Z_{M5}$, only $Z_{B^+}$ transforms under
  gauge transformations, so this is the relevant piece when we want to understand
  instanton corrections to couplings of
  charged matter fields. The partition function $Z_{B^+}$ is essentially a theta function
  on the intermediate Jacobian ${\cal J}(M5)$ of the $M5$-brane, and is characterized by
  its vanishing locus (the theta divisor). Note that there is no justification for adding extra degrees
  of freedom at the intersection of the $M5$-instanton with the discriminant locus of the elliptic fibration.
  \item The structure of the various pieces in the partition function depends on the cohomology
  groups $H^k(M5)$ of the $M5$-brane and their Hodge structure. Sometimes even on perfectly smooth Calabi-Yau
  four-folds one has to deal with $M5$ worldvolumes that are not smooth, for example an $M5$-brane
  at the discriminant locus wrapping a nodal elliptic curve.
  When the worldvolume of the $M5$-brane is not smooth but has only normal
  crossing singularities, we should replace the conventional cohomology groups by their
  logarithmic version $H^k_{\rm log}(M5)$, where the logarithmic structure is the canonical one associated
  to the normal crossing singularities.
  \item The $G$-flux may induce a tadpole for $B^+$, causing the partition function to vanish.
  This is often (but not exclusively) related to the existence of a gauged $U(1)$ symmetry
  in the $4d$ effective theory.
  \item When this happens, we can cancel the tadpole if we can find suitable Wilson surface operators
  of the form $W(\alpha) = \exp(i \int_\alpha B^+)$ to insert in the partition function. These operators
  generally transform non-trivially under gauge transformations. In the $4d$
  effective theory, the modified partition function with the extra insertions is interpreted as calculating instanton corrections
  to holomorphic couplings of $U(1)$ charged chiral fields that are forbidden in perturbation theory due
  to the $U(1)$ symmetry.
  \item We can compare $M5$-instantons to $D3$-instantons in perturbative IIb by taking a Sen limit, although
  this changes the pre-factor. In this limit the $M5$-brane splits into two pieces whose intersection is the
  IIb $D3$-instanton. We again used the physical gauge approach to treat the $D3$-instanton. This led to a
  partition function of the form $Z_{D3} = Z_\phi Z_\theta Z_F Z_{\lambda_{37}}$, which we can then compare with $Z_{M5}$.
  The logarithmic cohomology groups of the $M5$-brane worldvolume can be mapped to various cohomology
  groups on the $D3$-brane worldvolume and the $D3$-$D7$ intersection $\Sigma_{37}$.
  Qualitatively we can match all the pieces of $Z_{M5}$ to $Z_{D3}$.
  \item In the IIb limit, the partition function $Z_{B^+}$ reproduces both the partition function $Z_F$ for the $D3$ gauge field, as well
  as the partition function $Z_{\lambda_{37}}$ of chiral $D3$-$D7$ strings. Both of these are essentially theta functions on generalizations
  of Jacobian varieties, not the Jacobians themselves due to compatibility with the orientifold symmetry (in the Riemann surface case it is called a Prym). We used
  the conventional fermionic representation of $Z_{\lambda_{37}}$, although by the $2d$ Bose-Fermi equivalence it also admits a bosonized description which is more appropriate for comparing with the chiral two-form $B^+$.
  A tadpole for $Z_{B^+}$ corresponds to a tadpole for $Z_F$ or the presence of fermionic zero modes for $Z_{\lambda_{37}}$. The insertion
  of Wilson surface operators $W(\alpha)$ on the $M5$-brane corresponds to insertion of $\lambda_{37}$ zero modes in $Z_{\lambda_{37}}$,
  or Wilson operators in $Z_F$.
  \item If the $F$-theory model has a K3-fibration, then the IIb model has a type I dual. In this case
  it is apparent from \cite{Clingher:2012rg} that $Z_{\lambda_{37}}$ corresponds to the partition function $Z_{\lambda_{19}}$ of $D1$-$D9$ strings
  in the type I dual,
  or to the partition function $Z_{\lambda_{SO(32)}}$ of the fermionic
  left-movers in the $SO(32)$ heterotic dual.
  \item The $D3$ instanton worldvolume may itself be reducible. This leads to the subject of gluing branes,
  which we elaborated on in \cite{Donagi:2011jy,Donagi:2011dv}.
  \item Semi-realistic models in perturbative IIb or the $SO(32)$ heterotic string
  contain a $U(n)$ gauge group, rather than an $SU(n)$ gauge group.
  As we have discussed elsewhere \cite{Donagi:2009ra,Pantev:2009de}, the extra $U(1)\subset U(n)$ is absent in generic $F$-theory models, $M$-theory models
  or in the $E_8\times E_8$ heterotic string, in the sense that its mass is not parametrically small compared to the KK scale. This is one of the main qualitative
  differences between semi-realistic models in perturbative IIb and generic $F$-theory models.
  \item The extra $U(1) \subset U(n)$ imposes some rather strong constraints on the $4d$ effective theory.
  Perturbatively it forces the vanishing
  of top quark Yukawa couplings in $SU(5)$ GUT models. Non-perturbatively (in $1/m_{10}R$) it has been argued that
  this $U(1)$ symmetry eliminates certain instanton
  contributions to the superpotential that are used in moduli stabilization scenarios \cite{Blumenhagen:2007sm}. In generic $F$-theory
  models however (or in the $E_8\times E_8$ heterotic string), there is no such $U(1)$ symmetry and these constraints are absent.
  \item We may further compare $M5$-instantons with (half)-heterotic models, by taking the global-to-local limit of \cite{Donagi:2012ts}.
  The $M5$-instanton again splits into two pieces, and the piece that talks to the visible sector wraps a $dP_9$ surface.
  This led to the concept of an MSW instanton, which is somewhat like a heterotic instanton but has a more general worldvolume
  theory. Again $Z_{B^+}$ factorizes in this limit, and we get a piece $Z_{\lambda}$ which is analogous to the partition function
  of the left-movers in the heterotic string, except it couples to a single $E_8$ space-time gauge symmetry.
  \item The partition function $Z_\lambda$ is again
  a theta function, and can be interpreted as the theta function of a Prym appearing in the spectral cover description of the
  $dP_9$ fibration. We get vanishing behaviour and interplay with gauged $U(1)$ symmetries just as in $F$-theory or type IIb.
  We also used the $D$-instanton approach to show how the instanton contributes to holomorphic couplings, by calculating derivatives
  with respect to the background fields. For example we obtained the ${\bf 27}^3$ coupling in $E_6$ GUT models or an $Xe^{-T}$ coupling used in some supersymmetry breaking scenarios. By holomorphy, such couplings survive away from the global-to-local limit.
  \item The computation of instanton contributions to holomorphic couplings is generally polluted by Christoffel symbols
  (coming from contact terms).
  This means that the instanton generically contributes to all possible holomorphic couplings,
   but unless we have more than the minimal number of zero modes, this is hard to demonstrate and the only natural quantity to compute
  is the contribution to the expected value of the superpotential itself.
   \item One can do very explicit algebraic computations of the moduli dependence of $Z_{B^+}$ in the global-to-local limit, by calculating
  the pull-back of the theta divisor to the compactification moduli space following \cite{Buchbinder:2002ic}. One could use the very same strategy in the IIB limit.
  This should have interesting applications to moduli stabilization. But the problem of summing contributions of different
  instantons in the same homology class, if such exist, is not well understood in the physical gauge approach. In fact in certain situations it has been argued that the contributions can cancel due to a residue theorem \cite{Beasley:2005iu}.
\end{enumerate}

\bigskip

{\it Acknowledgements:} We would like to thank Chris Beasley for
initial collaboration and many fruitful discussions. We would
further like to thank J.~Guffin, S.~Katz, E.~Witten for discussion,
and C.~Vafa for generous support. Ron Donagi acknowledges partial
support by NSF grants 0908487 and 0636606.


\newpage

\begin{thebibliography}{99}

\bibitem{Becker:1995kb}
  K.~Becker, M.~Becker and A.~Strominger,
  ``Five-Branes, Membranes And Nonperturbative String Theory,''
  Nucl.\ Phys.\  B {\bf 456}, 130 (1995)
  [arXiv:hep-th/9507158].
  %%CITATION = NUPHA,B456,130;%%


\bibitem{Harvey:1999as}
  J.~A.~Harvey and G.~W.~Moore,
  ``Superpotentials and membrane instantons,''
  arXiv:hep-th/9907026.
  %%CITATION = HEP-TH/9907026;%%


\bibitem{Witten:1999eg}
  E.~Witten,
  ``World-sheet corrections via D-instantons,''
  JHEP {\bf 0002}, 030 (2000)
  [arXiv:hep-th/9907041].
  %%CITATION = JHEPA,0002,030;%%


\bibitem{Vafa:1996xn}
  C.~Vafa,
  ``Evidence for F-Theory,''
  Nucl.\ Phys.\  B {\bf 469}, 403 (1996)
  [arXiv:hep-th/9602022].
  %%CITATION = NUPHA,B469,403;%%

\bibitem{Witten:1996hc}
  E.~Witten,
  ``Five-brane effective action in M-theory,''
  J.\ Geom.\ Phys.\  {\bf 22}, 103 (1997)
  [arXiv:hep-th/9610234].
  %%CITATION = JGPHE,22,103;%%

\bibitem{Donagi:2008ca}
  R.~Donagi and M.~Wijnholt,
  ``Model Building with F-Theory,''
  arXiv:0802.2969 [hep-th].
  %%CITATION = ARXIV:0802.2969;%%


\bibitem{Beasley:2008dc}
  C.~Beasley, J.~J.~Heckman and C.~Vafa,
  ``GUTs and Exceptional Branes in F-theory - I,''
  JHEP {\bf 0901}, 058 (2009)
  [arXiv:0802.3391 [hep-th]].
  %%CITATION = JHEPA,0901,058;%%

\bibitem{Hayashi:2008ba}
  H.~Hayashi, R.~Tatar, Y.~Toda, T.~Watari and M.~Yamazaki,
  ``New Aspects of Heterotic--F Theory Duality,''
  Nucl.\ Phys.\  B {\bf 806}, 224 (2009)
  [arXiv:0805.1057 [hep-th]].
  %%CITATION = NUPHA,B806,224;%%


\bibitem{Sen:1997gv}
  A.~Sen,
  ``Orientifold limit of F-theory vacua,''
  Phys.\ Rev.\  D {\bf 55}, 7345 (1997)
  [arXiv:hep-th/9702165].
  %%CITATION = PHRVA,D55,7345;%%

\bibitem{Clingher:2012rg}
  A.~Clingher, R.~Donagi and M.~Wijnholt,
  ``The Sen Limit,''
  arXiv:1212.4505 [hep-th].
  %%CITATION = ARXIV:1212.4505;%%


\bibitem{Ganor:1996pe}
  O.~J.~Ganor,
  ``A note on zeroes of superpotentials in F-theory,''
  Nucl.\ Phys.\  B {\bf 499}, 55 (1997)
  [arXiv:hep-th/9612077].
  %%CITATION = NUPHA,B499,55;%%

%\cite{Pantev:2009de}
\bibitem{Donagi:2009ra}
  R.~Donagi and M.~Wijnholt,
  ``Higgs Bundles and UV Completion in F-Theory,''
  arXiv:0904.1218 [hep-th].
  %%CITATION = ARXIV:0904.1218;%%

\bibitem{Blumenhagen:2007sm}
  R.~Blumenhagen, S.~Moster and E.~Plauschinn,
  ``Moduli Stabilisation versus Chirality for MSSM like Type IIB
  Orientifolds,''
  JHEP {\bf 0801}, 058 (2008)
  [arXiv:0711.3389 [hep-th]].
  %%CITATION = JHEPA,0801,058;%%






\bibitem{Donagi:2012ts}
  R.~Donagi, S.~Katz and M.~Wijnholt,
  ``Weak Coupling, Degeneration and Log Calabi-Yau Spaces,''
  arXiv:1212.0553 [hep-th].
  %%CITATION = ARXIV:1212.0553;%%


\bibitem{Maldacena:1997de}
  J.~M.~Maldacena, A.~Strominger and E.~Witten,
  ``Black hole entropy in M-theory,''
  JHEP {\bf 9712}, 002 (1997)
  [arXiv:hep-th/9711053].
  %%CITATION = JHEPA,9712,002;%%


\bibitem{Buchbinder:2002ic}
  E.~I.~Buchbinder, R.~Donagi and B.~A.~Ovrut,
  ``Superpotentials for vector bundle moduli,''
  Nucl.\ Phys.\  B {\bf 653}, 400 (2003)
  [arXiv:hep-th/0205190].
  %%CITATION = NUPHA,B653,400;%%


\bibitem{Donagi:2011jy}
  R.~Donagi and M.~Wijnholt,
  ``Gluing Branes, I,''
  arXiv:1104.2610 [hep-th].
  %%CITATION = ARXIV:1104.2610;%%

\bibitem{Donagi:2011dv}
  R.~Donagi and M.~Wijnholt,
  ``Gluing Branes II: Flavour Physics and String Duality,''
  arXiv:1112.4854 [hep-th].
  %%CITATION = ARXIV:1112.4854;%%


\bibitem{Blumenhagen:2009qh}
  R.~Blumenhagen, M.~Cvetic, S.~Kachru and T.~Weigand,
  ``{\small D}-Brane Instantons in Type {II} Orientifolds,''
  Ann.\ Rev.\ Nucl.\ Part.\ Sci.\  {\bf 59}, 269 (2009)
  [arXiv:0902.3251 [hep-th]].
  %%CITATION = ARNUA,59,269;%%

\bibitem{Heckman:2008es}
  J.~J.~Heckman, J.~Marsano, N.~Saulina, S.~Schafer-Nameki and C.~Vafa,
  ``Instantons and SUSY breaking in F-theory,''
  arXiv:0808.1286 [hep-th].
  %%CITATION = ARXIV:0808.1286;%%

\bibitem{Cvetic:2009ah}
  M.~Cvetic, I.~Garcia-Etxebarria and R.~Richter,
  ``Branes and instantons at angles and the F-theory lift of O(1) instantons,''
  AIP Conf.\ Proc.\  {\bf 1200}, 246 (2010)
  [arXiv:0911.0012 [hep-th]].
  %%CITATION = APCPC,1200,246;%%

\bibitem{Cvetic:2010rq}
  M.~Cvetic, I.~Garcia-Etxebarria and J.~Halverson,
  ``Global F-theory Models: Instantons and Gauge Dynamics,''
  arXiv:1003.5337 [hep-th].
  %%CITATION = ARXIV:1003.5337;%%

\bibitem{Blumenhagen:2010ja}
  R.~Blumenhagen, A.~Collinucci and B.~Jurke,
  ``On Instanton Effects in F-theory,''
  arXiv:1002.1894 [hep-th].
  %%CITATION = ARXIV:1002.1894;%%

\bibitem{Munich}
``MSW Quantum Cohomology,'' talk delivered at the workshop ``GUTs
and Strings'' in Munich on February 12, 2010.%%
\newline
http://wwwth.mppmu.mpg.de/members/blumenha/gutworkshop/index.html.


\bibitem{Beasley:2005iu}
  C.~Beasley and E.~Witten,
  ``New instanton effects in string theory,''
  JHEP {\bf 0602}, 060 (2006)
  [arXiv:hep-th/0512039].
  %%CITATION = JHEPA,0602,060;%%

\bibitem{Wess:1992cp}
  J.~Wess and J.~Bagger,
  ``Supersymmetry and supergravity,''
%\href{http://www.slac.stanford.edu/spires/find/hep/www?irn=5426545}{SPIRES entry}
{\it  Princeton, USA: Univ. Pr. (1992) 259 p}

\bibitem{Curio:2009wn}
  G.~Curio,
  ``Perspectives on Pfaffians of Heterotic World-sheet Instantons,''
  arXiv:0904.2738 [hep-th].
  %%CITATION = ARXIV:0904.2738;%%



\bibitem{Distler:1988ms}
  J.~Distler and B.~R.~Greene,
  ``Some Exact Results on the Superpotential from Calabi-Yau
  Compactifications,''
  Nucl.\ Phys.\  B {\bf 309}, 295 (1988).
  %%CITATION = NUPHA,B309,295;%%

\bibitem{BGS3}
 J.-~M.~Bismut, H.~Gillet and C.~Soul\'e,
 {\it ``Analytic Torsion and Holomorphic Determinant Bundles III: Quillen
 Metrics on Holomorphic Determinants,''}
Comm.\ Math.\ Phys.\ {\bf 115}, 302-315 (1988).


\bibitem{AlvarezGaume:1986es}
  L.~Alvarez-Gaume, G.~W.~Moore and C.~Vafa,
  ``Theta functions, modular invariance, and strings,''
  Commun.\ Math.\ Phys.\  {\bf 106}, 1 (1986).
  %%CITATION = CMPHA,106,1;%%



\bibitem{Weinberg:2000cr}
  S.~Weinberg,
  ``The quantum theory of fields.  Vol. 3: Supersymmetry,''
%\href{http://www.slac.stanford.edu/spires/find/hep/www?irn=4384008}{SPIRES entry}
{\it  Cambridge, UK: Univ. Pr. (2000) 419 p}

\bibitem{Witten:1982hu}
  E.~Witten and J.~Bagger,
  ``Quantization Of Newton's Constant In Certain Supergravity Theories,''
  Phys.\ Lett.\  B {\bf 115}, 202 (1982).
  %%CITATION = PHLTA,B115,202;%%

\bibitem{Berglund:1995yu}
  P.~Berglund, P.~Candelas, X.~de la Ossa, E.~Derrick, J.~Distler and T.~Hubsch,
  ``On The Instanton Contributions To The Masses And Couplings Of E(6)
  Singlets,''
  Nucl.\ Phys.\  B {\bf 454}, 127 (1995)
  [arXiv:hep-th/9505164].
  %%CITATION = NUPHA,B454,127;%%

\bibitem{Dine:1987bq}
  M.~Dine, N.~Seiberg, X.~G.~Wen and E.~Witten,
  ``Nonperturbative Effects on the String World Sheet. 2,''
  Nucl.\ Phys.\  B {\bf 289}, 319 (1987).
  %%CITATION = NUPHA,B289,319;%%

\bibitem{Kallosh:2005gs}
  R.~Kallosh, A.~K.~Kashani-Poor and A.~Tomasiello,
  ``Counting fermionic zero modes on M5 with fluxes,''
  JHEP {\bf 0506}, 069 (2005)
  [arXiv:hep-th/0503138].
  %%CITATION = JHEPA,0506,069;%%

\bibitem{Saulina:2005ve}
  N.~Saulina,
  ``Topological constraints on stabilized flux vacua,''
  Nucl.\ Phys.\  B {\bf 720}, 203 (2005)
  [arXiv:hep-th/0503125].
  %%CITATION = NUPHA,B720,203;%%

\bibitem{Witten:1996bn}
  E.~Witten,
  ``Non-Perturbative Superpotentials In String Theory,''
  Nucl.\ Phys.\  B {\bf 474}, 343 (1996)
  [arXiv:hep-th/9604030].
  %%CITATION = NUPHA,B474,343;%%

\bibitem{Katz:1996th}
  S.~H.~Katz and C.~Vafa,
  ``Geometric engineering of N = 1 quantum field theories,''
  Nucl.\ Phys.\  B {\bf 497}, 196 (1997)
  [arXiv:hep-th/9611090].
  %%CITATION = NUPHA,B497,196;%%

\bibitem{Polyakov:1976fu}
  A.~M.~Polyakov,
  ``Quark Confinement And Topology Of Gauge Groups,''
  Nucl.\ Phys.\  B {\bf 120}, 429 (1977).
  %%CITATION = NUPHA,B120,429;%%

\bibitem{Diaconescu:1998ua}
  D.~E.~Diaconescu and S.~Gukov,
  ``Three dimensional N = 2 gauge theories and degenerations of Calabi-Yau
  four-folds,''
  Nucl.\ Phys.\  B {\bf 535}, 171 (1998)
  [arXiv:hep-th/9804059].
  %%CITATION = NUPHA,B535,171;%%


\bibitem{deBoer:1997kr}
  J.~de Boer, K.~Hori and Y.~Oz,
  ``Dynamics of N = 2 supersymmetric gauge theories in three dimensions,''
  Nucl.\ Phys.\  B {\bf 500}, 163 (1997)
  [arXiv:hep-th/9703100].
  %%CITATION = NUPHA,B500,163;%%

\bibitem{Witten:1999vg}
  E.~Witten,
  ``Duality relations among topological effects in string theory,''
  JHEP {\bf 0005}, 031 (2000)
  [arXiv:hep-th/9912086].
  %%CITATION = JHEPA,0005,031;%%

\bibitem{Henningson:1999dm}
  M.~Henningson, B.~E.~W.~Nilsson and P.~Salomonson,
  ``Holomorphic factorization of correlation functions in  (4k+2)-dimensional
  (2k)-form gauge theory,''
  JHEP {\bf 9909}, 008 (1999)
  [arXiv:hep-th/9908107].
  %%CITATION = JHEPA,9909,008;%%


\bibitem{Donagi:2008kj}
  R.~Donagi and M.~Wijnholt,
  ``Breaking GUT Groups in F-Theory,''
  arXiv:0808.2223 [hep-th].
  %%CITATION = ARXIV:0808.2223;%%

%\cite{Donagi:2008ca}
\bibitem{Curio:1998bva}
  G.~Curio and R.~Y.~Donagi,
  ``Moduli in N = 1 heterotic/F-theory duality,''
  Nucl.\ Phys.\  B {\bf 518}, 603 (1998)
  [arXiv:hep-th/9801057].
  %%CITATION = NUPHA,B518,603;%%

\bibitem{Polchinski:1998rr}
  J.~Polchinski,
  ``String theory. Vol. 2: Superstring theory and beyond,''
%\href{http://www.slac.stanford.edu/spires/find/hep/www?irn=4634802}{SPIRES entry}
{\it  Cambridge, UK: Univ. Pr. (1998) 531 p}


\bibitem{Pantev:2009de}
  T.~Pantev and M.~Wijnholt,
  ``Hitchin's Equations and M-Theory Phenomenology,''
  arXiv:0905.1968 [hep-th].
  %%CITATION = ARXIV:0905.1968;%%


%\cite{Donagi:2008kj}
\end{thebibliography}
\end{document}